\documentclass{aa}

\usepackage{graphicx}
\usepackage{color}
\definecolor{orange}{rgb}{0.8,0.4,0.0}
\usepackage[colorlinks=true,
citecolor=blue,
draft=false]{hyperref}
\usepackage[normalem]{ulem}
\usepackage{xspace}
\usepackage{txfonts}
\newcommand{\kmpers}{\ensuremath{\mathrm{km \, s^{-1}}}\xspace} 

\newcommand{\vlsr}{$\upsilon_{\rm LSR}$}              

\newcommand{\expo}[1]{$10^{#1}$}
\newcommand{\texpo}[1]{$\,\times\,10^{#1}$}

\newcommand{\ohtvaoett}{$\rm{H_2O\,(1_{10}-1_{01})}$}

\newcommand{\tolvco}{$^{12}$CO\xspace}
\newcommand{\trettenco}{$^{13}$CO\xspace}
\newcommand{\catteno}{C$^{18}$O\xspace}

\newcommand{\adeg}{$^{\circ}$}

\newcommand{\atwozero}{$\alpha_{2000}$}
\newcommand{\dtwozero}{$\delta_{2000}$}

\newcommand{\radot}[4]{\mbox{#1$^{\rm h}$#2$^{\rm m}$#3$\stackrel{^{\rm
s}}{_{\bf\cdot}}$#4}}
\newcommand{\decdot}[3]{\mbox{#1$^{\circ}$#2$^{\prime}$#3$^{\prime \prime}$}}
\newcommand{\msun}{\ensuremath{M_{\odot}}\xspace}

\authorrunning{Per Bjerkeli, et al.}
\titlerunning{ALMA observations of au scale kinematics in B335}

\begin{document} 

  \title{Kinematics around the B335 protostar down to au scales}

   \author{Per Bjerkeli \inst{1}
          \and Jon P. Ramsey \inst{2,4}
          \and Daniel Harsono \inst{3}
          \and Hannah Calcutt \inst{1}
          \and Lars E. Kristensen \inst{4} \and \\
          Matthijs H. D. van der Wiel \inst{5}
          \and Jes K. J{\o}rgensen \inst{4}
          \and S\'ebastien Muller \inst{1}
           \and Magnus V. Persson \inst{1}
          }

   \institute{Department of Space, Earth, and Environment,
              Chalmers University of Technology,\\
              Onsala Space Observatory, 439 92 Onsala, Sweden\\
              \email{per.bjerkeli@chalmers.se}
              \and
              Department of Astronomy, University of Virginia, Charlottesville, VA 22904, USA
              \and
              Leiden Observatory, Leiden University, P.O. Box 9513, NL-2300 RA Leiden, The Netherlands
              \and
              Niels Bohr Institute and Centre for Star and Planet Formation, University of Copenhagen, {\O}ster Voldgade 5--7,
DK-1350 Copenhagen K, Denmark
              \and
              ASTRON, the Netherlands Institute for Radio Astronomy, Oude Hoogeveensedijk 4, 7991 PD Dwingeloo, The
Netherlands\\
             }

   \date{Submitted May 24, 2019; accepted September 06, 2019}


  \abstract
  {The relationship between outflow launching and the formation of accretion disks around young stellar objects is still not entirely understood, which is why spectrally and spatially resolved observations are needed. Recently, the Atacama Large Millimetre/sub-millimetre Array (ALMA) has carried out long-baseline observations towards a handful of young sources, revealing connections between outflows and the inner regions of disks.}
  {Here we aim to determine the small-scale kinematical and morphological properties of the outflow from the isolated protostar B335 for which no Keplerian disk has, so far, been observed on scales down to 10~au.}
  {We use ALMA in its longest-baseline configuration to observe emission from CO isotopologues, SiO, SO$_2$ and CH$_3$OH. The proximity of B335 provides a resolution of $\sim$3~au (0.03\arcsec). We also combine our long-baseline data with archival observations to produce a high-fidelity image covering scales up to 700~au (7\arcsec).}
  {\tolvco has a X-shaped morphology with arms $\sim$50~au in width that we associate with the walls of an outflow cavity, similar to what is observed on larger scales. Long-baseline continuum emission is confined to $<$7~au from the protostar, while short-baseline continuum emission follows the \tolvco\ outflow and cavity walls. Methanol is detected within $\sim$30~au of the protostar. SiO is also detected in the vicinity of the protostar, but extended along the outflow.}
  {The \tolvco outflow shows no clear signs of rotation at distances $\gtrsim$30~au from the protostar. SiO traces the protostellar jet on small scales, but without obvious rotation. CH$_3$OH and SO$_2$ trace a region $<$16~au in diameter, centred on the continuum peak, which is clearly rotating. Using episodic, high-velocity, \tolvco\ features, we estimate the launching radius of the outflow to be $<$0.1~au and dynamical timescales on the order of a few years.}

\maketitle  
\section{Introduction}
\label{sec:intro}
Protostellar outflows are arguably the most visibly prominent signature of ongoing star formation. Since their discovery almost four decades ago \citep{Snell:1980lr}, a range of models have been proposed to explain how they are launched. The differences between these models are mainly in where acceleration of the outflow takes place \citep[see, e.g.][for a recent review]{Frank:2014ys}, i.e., close to the protostar \citep{Shu:1994kx}, or throughout an extended region in the disk \citep{Blandford:1982fj,Pudritz:1983fv,Lynden-Bell:1996rk}. What all these models share in common, though, is the assertion that outflows are magnetically-powered. 

Recently, we reported on the first resolved images of an outflow launching region towards the Class~I source TMC1A \citep{Bjerkeli:2016wo}, demonstrating that launching is occurring from the disk at radii up to $\sim$25~au. Other observations carried out at high spatial resolution have since shown that launching can take place out to large radii in the disk \citep[e.g.][]{Lee:2018et,Alves:2017rt}, but also close to the protostar itself \citep[e.g. HH\,212;][]{Lee:2018et}. These results underscore that outflow launching itself is still not entirely understood and, in particular, how it varies with the evolutionary stage of the disk.

To understand the launching of outflows, and specifically how angular momentum is transported throughout the system, one first needs to comprehend disk formation and its relation to outflow launching. Disk formation begins when a rotating molecular cloud core contracts under its own gravity \citep[e.g.][]{Shu:1987fk}. Due to the conservation of angular momentum, during this inside-out collapse, infalling material which has too much angular momentum to fall directly onto the central protostar instead forms a rotating disk. The presence of a magnetic field during collapse, meanwhile, provides a means to efficiently transport angular momentum, which can strongly affect disk formation. The initial conditions for the disk structure, as well as its early evolution, are set during this deeply embedded phase of star formation, i.e., the Class 0 stage \citep{Andre:2000qy}. In this phase, collapse proceeds inside-out, and infalling motion initially dominates over rotational motions. Over time, the infalling material begins to accumulate in a disk that slowly settles into nearly Keplerian rotation. The details of when and how the disk becomes Keplerian, however, is still under debate \citep[e.g.][]{Li:2014jm, Wurster:2018re}.

In the presence of a large-scale magnetic field, the interaction between the rotation of the system (disk + star) and the magnetic field promotes the launching of an outflow perpendicular to the disk/star rotation axis. This outflow carries with it not only angular momentum extracted from the disk, but also mass. Numerical studies have, meanwhile, shown that outflows can form already during first-core formation \citep{Tomida:2013yg}.  Numerous studies \citep[e.g.][]{Li:2011tt,Machida:2014lr,Tomida:2015yu} have shown that magnetic fields can also slow down infall and prevent the formation of a Keplerian  disk (the so-called ``magnetic braking catastrophe'') during the earliest stages. In this scenario, in order to form disks, one must introduce non-ideal magnetic effects \citep[e.g.][]{Zhao:2016wy} and/or turbulence \citep[e.g.][]{Seifried:2013dc,Li:2014km}.

The advent of the Atacama Large Millimetre/sub-millimetre Array (ALMA) is revolutionising our understanding of star and disk formation. To date, ALMA observations have revealed Keplerian disks towards several very young sources \citep[e.g.][]{Tobin:2012fr,Murillo:2013fz,Lindberg:2014zr,Ohashi:2014fe}. However, ALMA has also revealed counter examples where a Keplerian disk has not yet been detected. One such example is the Class~0 protostar B335.

B335 is an isolated dense globule associated with the infrared identified protostar ``IRAS 19347+0727'' at a distance of approximately 100 pc \citep[90--120 pc;][]{Olofsson:2009jl}.  Although recent studies \citep{Evans:2015qp,Yen:2015vf} have revealed that matter is infalling towards the centre of B335, these studies were not able to resolve any Keplerian component on scales greater than $\sim$10~au in size. From the standard paradigm of protostellar disk formation theory, the absence of a rotationally supported disk on these scales suggests that the B335 system is either very young, and/or that it is subject to strong magnetic braking \citep[e.g.][]{Yen:2015vf}.

A young age for B335 is also supported by recent single-dish observations of its outflow. Observations of the molecular component \citep{Yildiz:2015kq} imply that dynamical time-scales for the CO emitting gas are of the order of \expo{4}~yr. This suggests that the B335 system is extremely young, and therefore, it is an ideal target to study early and ongoing star and disk formation. Furthermore, B335 is known to power a fast protostellar jet and associated Herbig-Haro (HH) objects \citep{Galfalk:2007lr,Reipurth:1992rw}. Based on the proper motions of the HH objects, the estimated dynamical time-scale is only a few hundred years.

Several saturated complex organic molecules indicative of hot corino chemistry (e.g.\ CH$_3$CHO, HCOOCH$_3$ and NH$_2$CHO) have been detected on small scales towards B335 \citep{Imai:2016lr}. More recently, \citet{Imai:2019ip} observed rotation in CH$_3$OH and HCOOH towards B335 on scales of $\sim$0.2\arcsec ($\sim$20~au). CH$_3$OH is a molecule of key interest when it comes to the formation of organics \citep{Herbst:2009uo}. Understanding the link between emerging disks and such complex species therefore remains very important for understanding the initial conditions of protoplanetary disk chemistry. While CH$_3$OH is often seen to be associated with high column density material in protostellar envelopes, as well as shocks in outflows, to date, it has only been detected in the disks around two sources, TW Hya and V883~Ori \citep{Walsh:2016qy,vant-Hoff:2018qk}, both of which are significantly older than B335.

The morphology of the molecular component \citep[e.g.][]{Yen:2010qf} traced by CO\,(2--1), meanwhile, reveals an X-shaped structure with an opening angle of the order of $\sim$45\adeg. The outflow is nearly in the plane of the sky with an inclination angle that is between 3\adeg\ and 10\adeg\ \citep{Hirano:1988lr,Stutz:2008fk}\footnote{Herein, we define an inclination angle of $0^\circ$ to coincide with features that lie in the plane of the sky.}.

The age, the orientation, and the proximity of B335 make it an excellent target for long-baseline observations with ALMA. In its most extended configuration, ALMA can attain an angular resolution of $\sim$0.02\arcsec, providing a maximum linear resolution of $\sim$2~au for the distance of B335. Such a high spatial resolution enables detailed studies of both the innermost protostellar region and the outflow launching region.
\section{Observations}
\label{sec:observations}
\subsection{New long-baseline observations}
\label{sub:longbaseline}
B335 was observed with ALMA between October 21st and 29th 2017 (five execution blocks) as part of the Cycle 5 program 2017.1.00288.S. Observations were carried out in Band 6 and cover five spectral windows (SPWs). Three of them were centred on CO isotopologues, viz., the $J$~=~2--1 transitions of $^{12}$CO, $^{13}$CO, and C$^{18}$O at 230.5~GHz, 220.4~GHz, and 219.6~GHz, respectively. The \mbox{SiO\,(5--4)} transition was covered in the fourth SPW centred at 217.1~GHz, while the fifth SPW targeted the continuum at 233.0~GHz. Between 45 and 51 antennas of the 12~m array were used. Baselines were in the range 41 -- 16196 m, yielding a spatial resolution of 0.03\arcsec and a largest recoverable scale corresponding to $\sim$0.3\arcsec\footnote{ALMA Technical Handbook, \url{https://almascience.eso.org}}. For the adopted distance of B335, this implies a linear resolution of $\sim$3~au, but also that these observations are not sensitive to structures with angular scales larger than $\sim$30~au, e.g., the foreground envelope emission and large-scale outflow emission. The spectral resolution was set to 122 kHz for the \tolvco SPW, and 61~kHz for \trettenco and \catteno, while the total bandwidth in each SPW was 120~MHz. The continuum was observed at a spectral resolution of 31.25 MHz, while the 217.1~GHz SPW centred on SiO had a bandwidth of 937~MHz and a spectral resolution 488~kHz. Over the course of the observations, the precipitable water vapour (PWV) ranged from 0.5 to 1.5 mm. The phase centre of the observations were \atwozero~=~\radot{19}{37}{00}{890}, \dtwozero~=~\decdot{+07}{34}{09.60}. J2000--1748, J2134--0153 and J2148+0657 were used as bandpass calibrators, while J1938+0448 was used to calibrate the phase. The flux calibration was done using J2000$-$1748, J2134$-$0153, and J2148+0657. We estimate a flux accuracy of $\sim$10\%. Data reduction and imaging were carried out in CASA v5.1.1 \citep{McMullin:2007nr}. The continuum was subtracted in the $uv$ domain using only line-free channels. Calibrated visibilities were then transferred into the image domain using the CLEAN algorithm with Briggs weighting and a robust parameter of 0.5. All spectral line maps were imaged at the native spectral resolution.

\begin{table*}[t]
\flushleft
\caption{Summary of all ALMA data sets employed in this study.}
\begin{tabular}{lllllllll}\hline\hline
  \noalign{\smallskip}
Project &  Date & $N_{\rm ant}$ & $t_{\rm on}$ & PWV & $B_{\rm min}$ & $B_{\rm max}$  & DR & $\Delta$ (R.A.,Dec.) \\
        &       &           &(s)      & (mm)&  (m)       & (m)       &   & (mas) \\

  \noalign{\smallskip}
\hline

2013.1.00879.S & 02-Sep-2014 & 34 & 1463 &  1.0 & 32 & 1052 & 69 & -80, 95 \\

2016.1.01552.S & 21-Nov-2016 & 43 & 1851 & 2.1 & 15 & 704 & 67 & 32, -65 \\
               & 19-Mar-2017 & 40 & 685 & 1.6 & 15 & 287 & 34 & -11, 24 \\

2017.1.00288.S & 08-Oct-2017 & 51 & 2268 & 0.5 &  41 & 16196 & 99 & -19, -1 \\ 
               & 21-Oct-2017 & 49 & 2268& 1.4 &  41 & 16196 & 64 & -15, 0 \\ 
               & 22-Oct-2017 & 46 & 2268& 1.3 &  41 & 16196 & 62 & -13, 0 \\ 
               & 27-Oct-2017 & 47 & 2271& 0.7 & 135 & 14851 & 94 & -14, -3 \\ 
               & 29-Oct-2017 & 45 &2268 & 1.5 & 113 & 13894 & 37 & -15, -2 \\ 
 
 \hline
  \noalign{\smallskip}
\end{tabular}
\tablefoot{
 \tablefoottext{*}{$N_{\rm ant}$ = number of antennas in the array; $t_{\rm on}$ = on-source integration time; PWV = precipitable water vapor; $B_{\rm min}$ = shortest projected baseline; $B_{\rm max}$ = longest projected baseline; 
 DR = dynamic range of the continuum data; $\Delta$(R.A., Dec.) = astrometry offsets for the continuum peak, relative to the phase centre set at \atwozero~=~\radot{19}{37}{00}{890}, \dtwozero~=~\decdot{+07}{34}{09.60}}.}
\label{table:combined}
\end{table*}

\subsection{Combining with archived data}
\label{sub:combineddata}
The largest recoverable scales in our long-baseline data are much smaller than the known extent of the CO\,(2--1) emission in B335 \citep[e.g.][]{Yen:2010qf}. As such, short-spacing data is needed to capture any large-scale outflow emission filtered out by our long-baseline observations and make the connection to our higher-resolution data. Our goal is, therefore, to produce a high-as-possible fidelity, and simultaneously high-angular resolution, image of the emission in B335, covering all relevant scales.

To accomplish this, we combined our long-baseline data with two publicly-available ALMA data sets towards B335 from 2014 (2013.1.00879.S) and 2016/2017 (2016.1.01552.S). The spectral setup for both archival data sets covered the \tolvco\,(2--1) line, and calibration was performed using CASA v4.3.1 and v4.7.0, respectively. Care is needed when combining data taken with different array configurations (see Table \ref{table:combined}), including w.r.t.\ the flux accuracy, relative astrometry, and handling of the weights in each individual data set. Data combination was carried out in the following manner: First, we cleaned the continuum emission from different data sets individually, using the H\"ogbom deconvolution algorithm \citep{Hogbom:1974zl}, reprojecting all data to the same phase centre. We used a 2D-Gaussian fit to the peak position of the compact continuum emission in each data set to constrain their relative astrometry. After performing the 2D Gaussian fits and a visual inspection, relatively small offsets (although in a few cases slightly larger than the final beam size) between the different data sets were found, as indicated in Table \ref{table:combined}. We also checked the dynamic range of each data set by taking the ratio of the continuum peak to the noise level in the residuals of the cleaned product. The weights of each data set were normalised by dividing them by their average (i.e., all data sets received the same average weight level). We found that the dynamic ranges were consistent within a factor of three, and hence, we decided to not modify further the relative weightings. During the combination, we manually flagged (and consequently never used) some of the data because it was of bad quality.

\begin{figure}[ht]
   \flushleft
   \includegraphics[width=0.9\hsize]{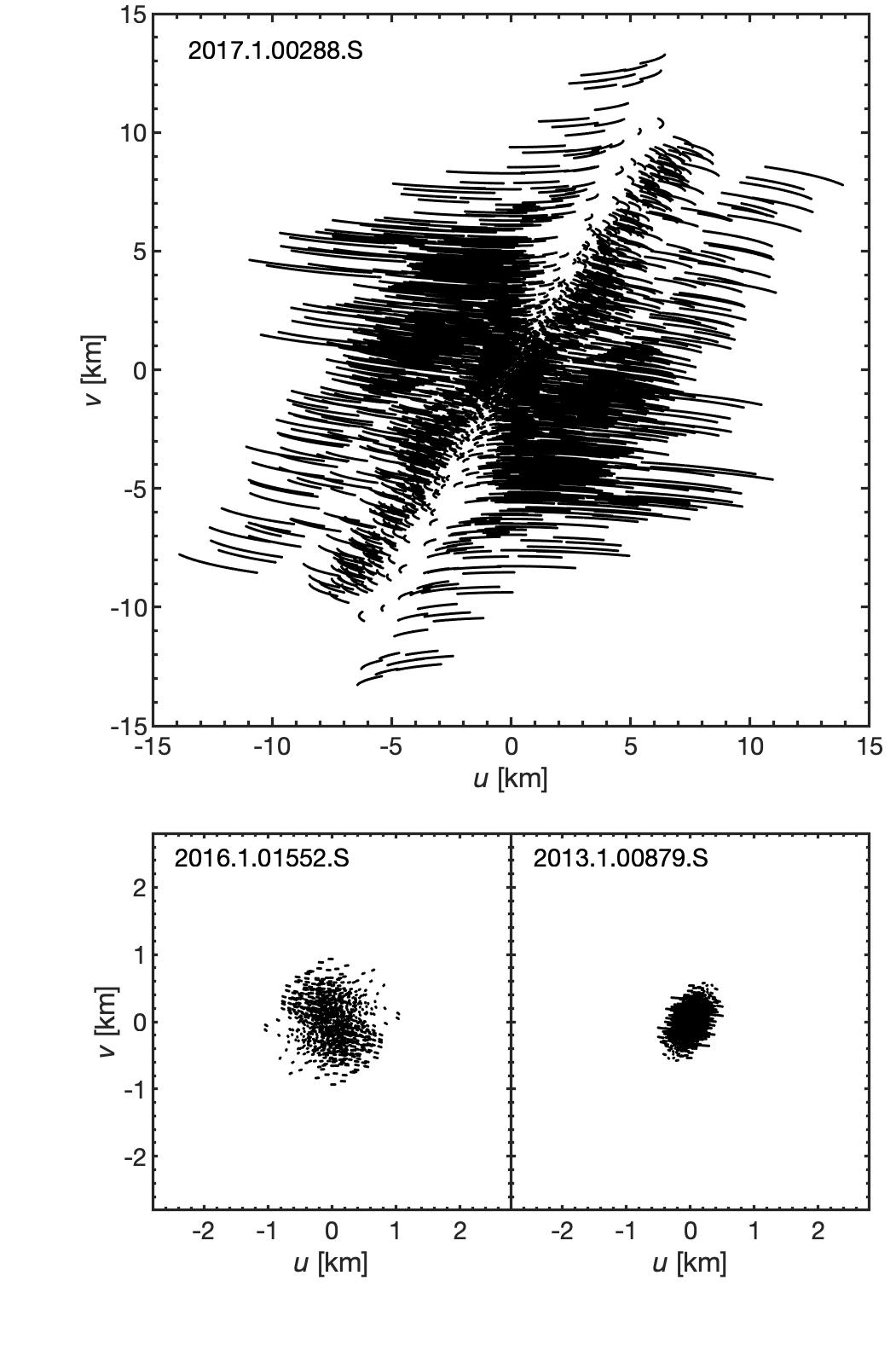}
      \caption{$u, v$ coverage for the three different ALMA projects. Note the change of scale between panels.}
  \label{fig:uvcoverage}
\end{figure}

Next, we deconvolved the combined data set using the CASA \texttt{tclean} procedure. Briggs weighting was used with the robust parameter set to 0.5. In all cases, the spectral resolution was sufficiently high to resolve the \tolvco\,(2--1) line with 0.25~\kmpers. The angular resolution of the 2013 project is $\sim$0.3\arcsec, while the 2016/2017 project was observed in two different configurations corresponding to spatial resolutions of $\sim$0.7\arcsec and $\sim$1.5\arcsec, respectively. By combining these data sets, we were able to recover emission on scales from $\sim$0.03\arcsec up to $\sim$7\arcsec (see Fig.\ \ref{fig:uvcoverage} for the different $u, v$ coverages). The cleaning threshold for the continuum was set to 0.05 mJy/beam and the threshold for the lines was set to 3 mJy/beam. Due to the nature of the combined data set, we used the so-called multiscale deconvolution algorithm for the line emission. The input scales used were 0, 8, 23, 68, 203, and 608 pixels, while the pixel scale was set to 0.01\arcsec. The size of the cleaned image is 4096 by 4096 pixels. For the line emission, we performed test cleans on a few channels close to the systemic velocity to confirm that changing the cleaning scales does not significantly affect the results. To ensure the combined image was consistent with the individual data sets, we also cleaned the different \tolvco\ data sets with the H\"ogbom deconvolution algorithm and compared the results to the multiscale cleaned image. Indeed, convolving the final combined data product with 2D Gaussians (with FWHM corresponding to the angular resolution of each data set) reproduces the overall morphology of the emission observed in projects 2013.1.00879 and 2016.1.01552.

It is worth noting that the multiscale cleaning of the combined data set required significant computing resources. For the combined \tolvco\ image presented in Fig.\ \ref{fig:moment0_combined}, one dedicated node (256 GB RAM) of the C3SE Hebbe cluster\footnote{Chalmers Centre for Computational Science and Engineering: \url{https://www.c3se.chalmers.se}} was used for $\sim$800 wall-clock hours. Even after that amount of time, a few channels were not properly cleaned (i.e.\ they had not reached the desired threshold of 3 mJy/beam).

\begin{figure}[ht]
   \flushleft
   \includegraphics[width=1.0\hsize]{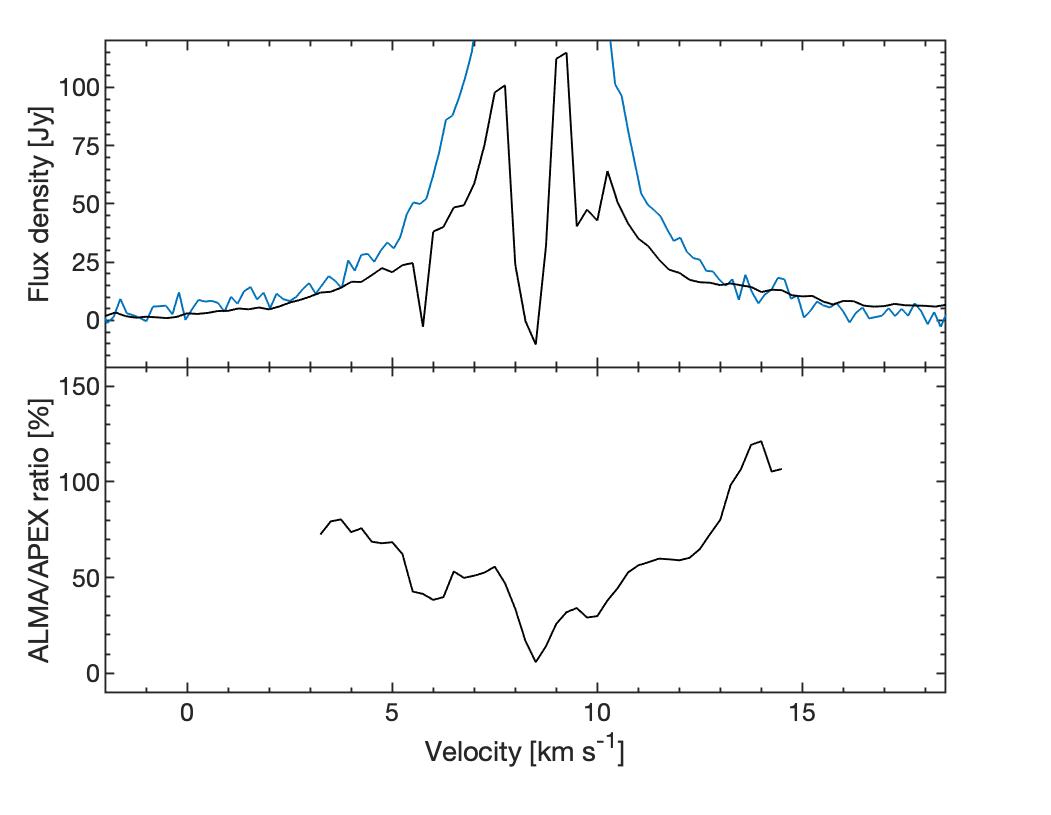}
      \caption{\textit{Upper panel:} Combined multiscale \tolvco spectrum (black), integrated over the field of view, as compared to the single-dish APEX spectrum (blue) observed towards the position of B335. \textit{Lower panel:} Moving average (using 1.0~\kmpers bins) of the ALMA-to-APEX ratio of fluxes in regions where the APEX S/N~$>$3$\sigma$. 
      } 
\label{fig:apexcomp}
\end{figure}

To check the amount of \tolvco~(2--1) flux recovered in these observations, we compared the final combined image with single dish data acquired with the Atacama Pathfinder Experiment (APEX) telescope (Project: O-087.F-9314A, PI: M.~V.~Persson). Fig.\ \ref{fig:apexcomp} demonstrates that more than 50\% of the emission observed with APEX is recovered in the line wings ($>$2\kmpers from systemic velocity) of the ALMA data. Also evident from this figure are the channels where cleaning was not carried out deeply enough, i.e., one channel at approximately +6.0~\kmpers and a few channels at +9.5 -- +10.0~\kmpers.

For the analysis that follows, we use the long-baseline data when analysing the \trettenco, \catteno, SiO, SO$_2$, and CH$_3$OH emission, while we use primarily the combined dataset when analysing the $^{12}$CO and continuum emission. The analysis was performed with MATLAB.

\section{Results}
\label{sec:results}
\subsection{Continuum and carbon monoxide}
\label{sub:continuumandcarbonmonoxide}
\begin{figure}
   \flushleft
   \includegraphics[width=1.0\hsize]{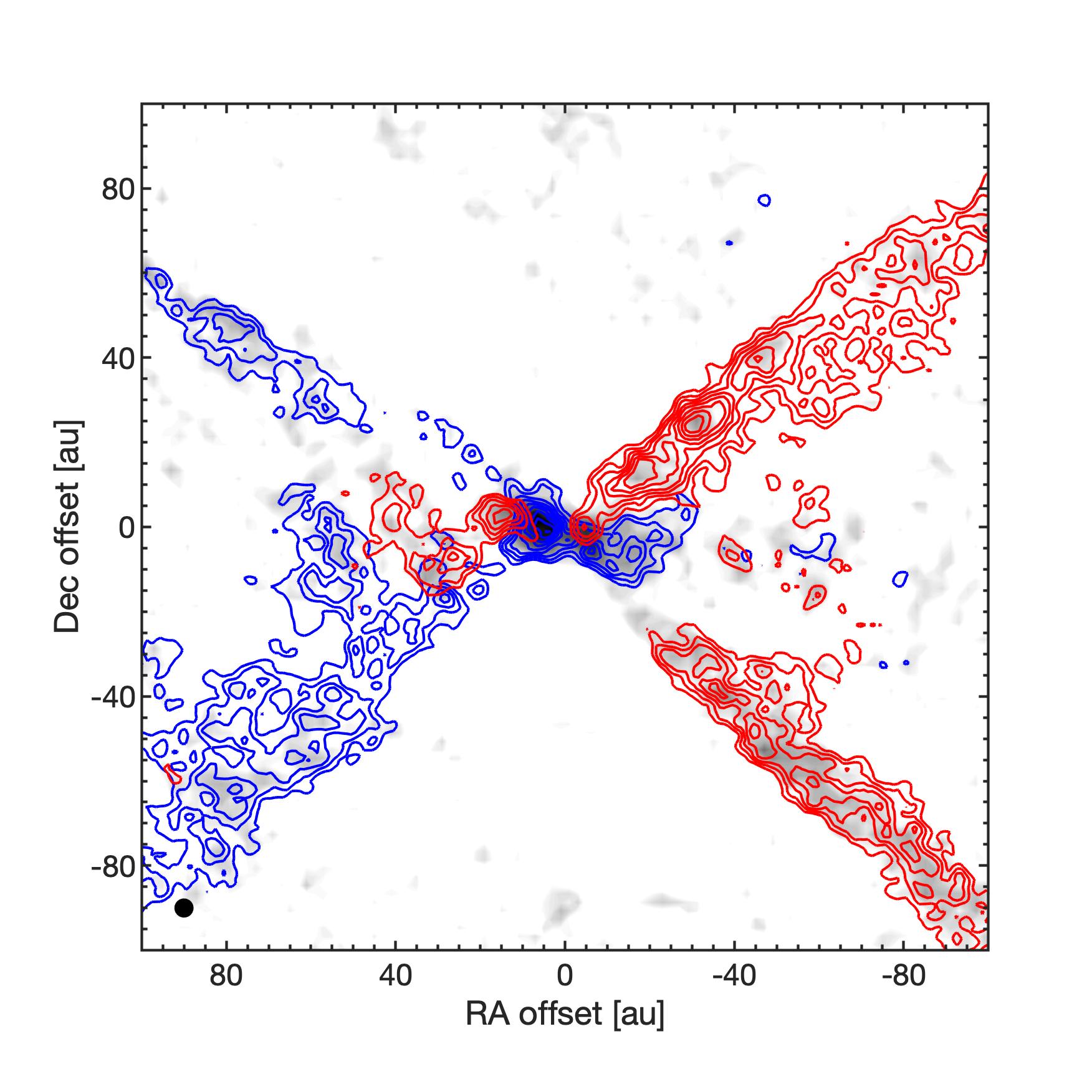}
   \caption{Combined \tolvco\,(2--1) emission convolved with a 2D Gaussian to a resolution of 4~au and integrated from 2.0 \kmpers to 10.0 \kmpers relative to the source velocity, 8.3 \kmpers\ w.r.t.\ \vlsr. Contours are from 3$\sigma$ in steps of 1$\sigma$, where $\sigma$~=~3.3~mJy beam$^{-1}$ \kmpers. Greyscale shows the corresponding map when using the long-baseline 2017.1.00288 data only. The convolved beam is shown in the lower left corner of the map.}
   \label{fig:lbvs288}
\end{figure}

\begin{figure*}
   \flushleft
   \includegraphics[width=1.0\hsize]{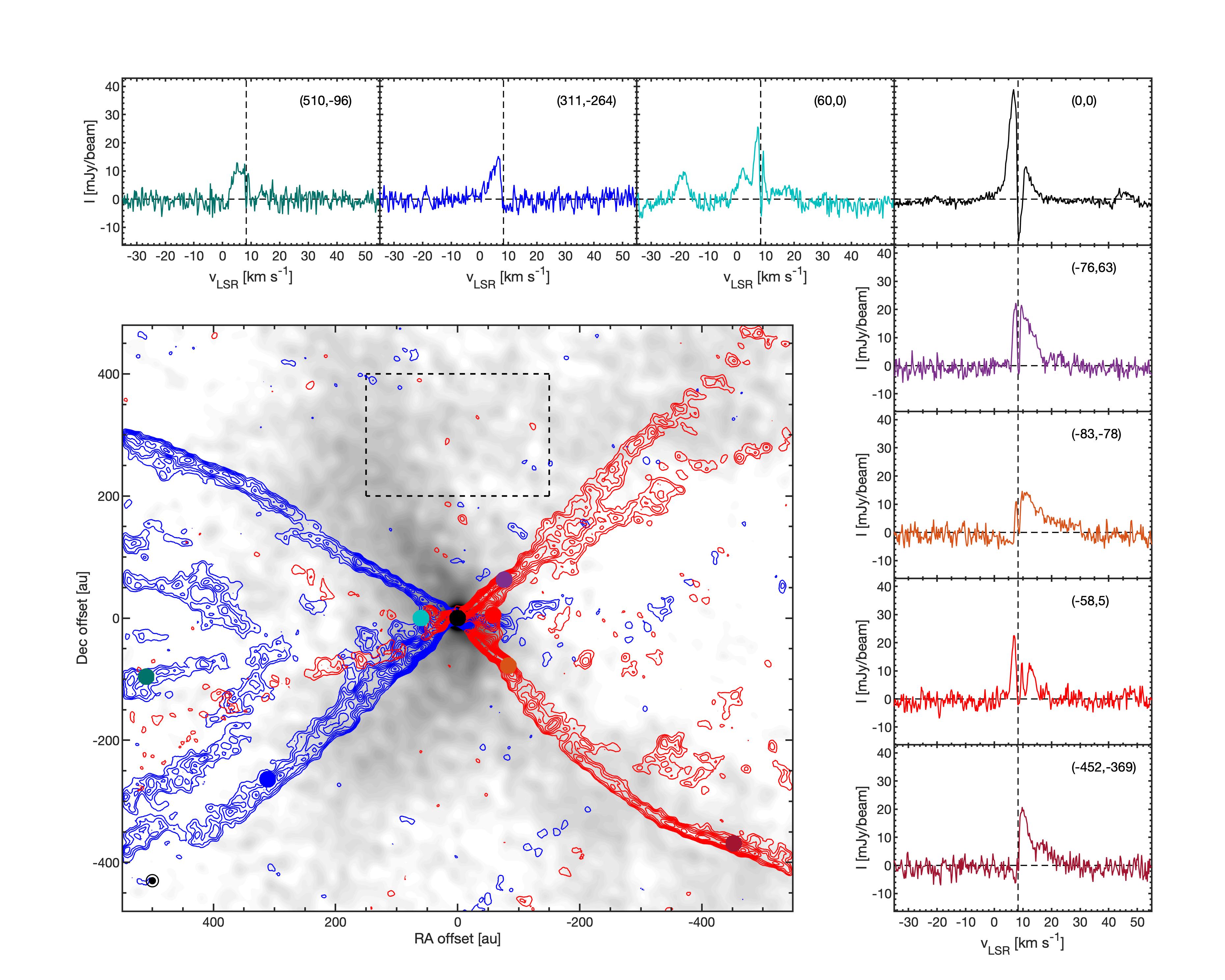}
   \caption{Moment 0 map of the \tolvco emission (contours from 3$\sigma$ in steps of 1$\sigma$, where $\sigma~=~$5.6~mJy~beam$^{-1}$), overlaid on the continuum (greyscale from 0 to 25\% of maximum), in the combined data set. The \tolvco image was convolved with a 2D Gaussian to 10~au resolution to enhance the S/N ratio in the map, while the continuum was convolved with a 2D Gaussian to 20~au resolution. The emission is integrated from 2.0 -- 6.0 \kmpers with respect to the source velocity, 8.3 \kmpers\ w.r.t. \vlsr. The dashed box denotes where the line rms was calculated using the line free channels. Selected mean spectra averaged over circular regions (10 au in radius) are indicated by the coloured points and the corresponding coloured spectral profiles. The convolved beams are shown in the lower left corner of the map. For clarity, a zoomed-out version of this figure with only the continuum data is presented in Fig.~\ref{fig:a0}.}
   \label{fig:moment0_combined}
\end{figure*}

In the long-baseline data, continuum emission at 1.3 mm was detected with a peak flux of 4.8~mJy/beam ($\sigma \simeq 2\times10^{-5}$~Jy) towards \atwozero~=~\radot{19}{37}{00}{900}, \dtwozero~=~\decdot{+07}{34}{09.52} (see Fig.\ \ref{fig:siovscont}). The emission is approximately Gaussian with a $\sim$5~au extension to the north-west. The full width half-maximum (FWHM) of the continuum was estimated using 2D Gaussian fitting in the image plane and found to be $\sim$7~au, i.e., slightly extended relative to the size of the synthesised beam. We also imaged the continuum using the combined data set. While the continuum is only detected towards the central position in the long-baseline observations, the combined data set shows extended emission on much larger scales (Figs.\ \ref{fig:moment0_combined} and \ref{fig:a0}).

The systemic velocity of B335 has previously been estimated to be in the 8.3 -- 8.5 ~\kmpers\ range \citep{Evans:2005qy,Jorgensen:2007fj,Yen:2011yu,Mottram:2014yq}. In the analysis presented in this paper, we adopt a source velocity of 8.3~\kmpers\ based on rare, isotopic line emission \citep{Evans:2005qy}. That is roughly consistent with what we find from the CO isotopologue observations presented here: By extracting mean spectra over a 50 au diameter circular region centred on the continuum peak, we find the systemic velocity to be $\sim$8.5~\kmpers\ w.r.t. \vlsr\ in all three CO isotopologues, independent of whether we use the combined data set or the long-baseline data only.

Previous observations \citep[e.g.][]{Yen:2010qf}, have shown that the CO\,(2--1) emission is prominent along the outflow cavity walls where envelope gas is expected to be entrained. That is consistent with what we find here. In our long-baseline data, \tolvco emission is detected predominantly towards the outflow emanating from B335 (peak flux $\sim$12~mJy beam$^{-1}$ at S/N $\simeq$ 7). From comparison with APEX single-dish data, however, it is obvious that a significant fraction of the emission originates on scales greater than the MRS of the long-baseline observations, viz., 0.3\arcsec\ (recovered flux is less than 20\% in the line wings). Therefore, for the remainder of the \tolvco analysis presented in this paper we use the multiscale cleaned image. Figure \ref{fig:lbvs288} presents the integrated \tolvco emission from the combined data set (contours) overlaid on top of the long-baseline only data (greyscale). A similar, but larger-scale and slightly smoothed ($\sim$0.1\arcsec) map is shown in Fig.\ \ref{fig:moment0_combined}, where spectra towards a few selected positions are also presented.

Maximum observed velocities of the main line component are $\sim$10~\kmpers w.r.t.\ the systemic velocity, consistent with previous single dish \tolvco\,(3--2) observations of B335 \citep{Yildiz:2015kq}. In addition, however, we observe two additional emission peaks at $-20$~\kmpers and at $+40$~\kmpers at positions close to the central source (see black point/spectra in Fig.\ \ref{fig:moment0_combined}) and in the outflow at a distance of $\sim$60~au from the central source (cyan point/spectra in Fig.\ \ref{fig:moment0_combined}), respectively. We interpret these peaks as higher velocity components potentially associated with the known high-velocity protostellar jet \citep{Galfalk:2007lr}. It should be noted, however, that the SiO\,(5--4) emission (Sect.\ \ref{sub:resultsSiOCH3OHSO2}) does not extend out to the positions where the high-velocity features are detected, but this may simply be due to insufficient sensitivity, and it is therefore difficult to confirm the origin of the high velocity \tolvco peaks. The observed velocity shifts of these components are, in fact, similar to the velocity extent of the \ohtvaoett\ line observed with Herschel-HIFI towards the central position \citep{Kristensen:2012kx,Mottram:2014yq}, but a more detailed comparison with these observations is hindered by the low signal-to-noise ratio of the water data.

The \tolvco emission shows a pronounced X-shaped morphology (Fig.\ \ref{fig:moment0_combined}) that first becomes apparent in our data set on scales $\sim$15~au, and extends well beyond the MRS of even the combined data to $\gtrsim$2000~au \citep[e.g.][]{Yen:2010qf}. A clear red-blue asymmetry where blueshifted emission is detected in the eastern outflow component, and redshifted emission is detected in the western outflow component, is observed. It is also evident that the emission is dominated by the cavity walls plus a moderate amount of emission predominantly in the blueshifted outflow between the cavity walls. We note the presence of a prominent secondary arc in the blueshifted component, offset $\gtrsim$300 au from the continuum peak, but, at present, its nature remains unclear. There is also a small amount of redshifted emission present on the blueshifted side of the outflow (and vice versa on the redshifted side). In the integrated emission map (Fig.\ \ref{fig:lbvs288}), one can also see a partial band of emission, roughly $\sim$50~au from the protostar, seemingly connecting the two blueshifted outflow cavity walls together. Traces of a counterpart in the redshifted outflow are also visible at the 3--4$\sigma$ level. These features are indeed spatially coincident with the aforementioned high velocity components. We will return to these features and their interpretation in Sect.~\ref{sec:discuss}.

\begin{figure}
   \centering
   \includegraphics[width=1.0\hsize,clip=true,trim=40 0 55 0]{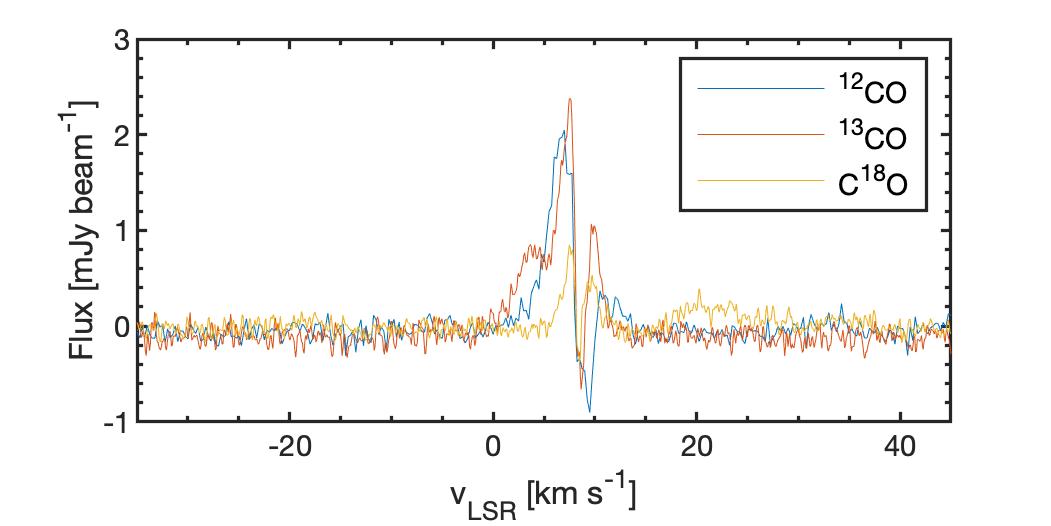} \\
   \caption{Mean $^{12}$CO, $^{13}$CO, and C$^{18}$O spectra extracted from a circular region of diameter 50 au centred on the protostar. Data is from the long-baseline observations only.}
  \label{figure:COspectra50au}
\end{figure}

In the long-baseline data set, \trettenco is detected only in close proximity to the central continuum peak. No significant \trettenco emission is detected towards the outflow cavity walls traced by \tolvco. The maximum observed velocities of \trettenco, however, coincide with the maximum velocities observed in \tolvco close to the central position. \catteno was also covered in our observations, but was only detected after degrading the spatial resolution, and even then only at low signal-to-noise. No conclusions can be drawn with regard to its spatial distribution and consequently we have chosen not to discuss the data further here. For completeness, and to aid comparison, Fig.\ \ref{figure:COspectra50au}, shows the spectra of the three CO isotopologues towards a 50 au sized region centred on the continuum peak. We note that all three line profiles show red-blue asymmetries indicating infall, and the line ratios suggest that the emission is optically thick in \tolvco and \trettenco.

\subsection{SiO, \texorpdfstring{CH$_3$OH}{CH3OH}, and \texorpdfstring{SO$_2$}{SO2}}
\label{sub:resultsSiOCH3OHSO2}
\begin{figure}
   \flushleft
   \includegraphics[width=1.0\hsize]{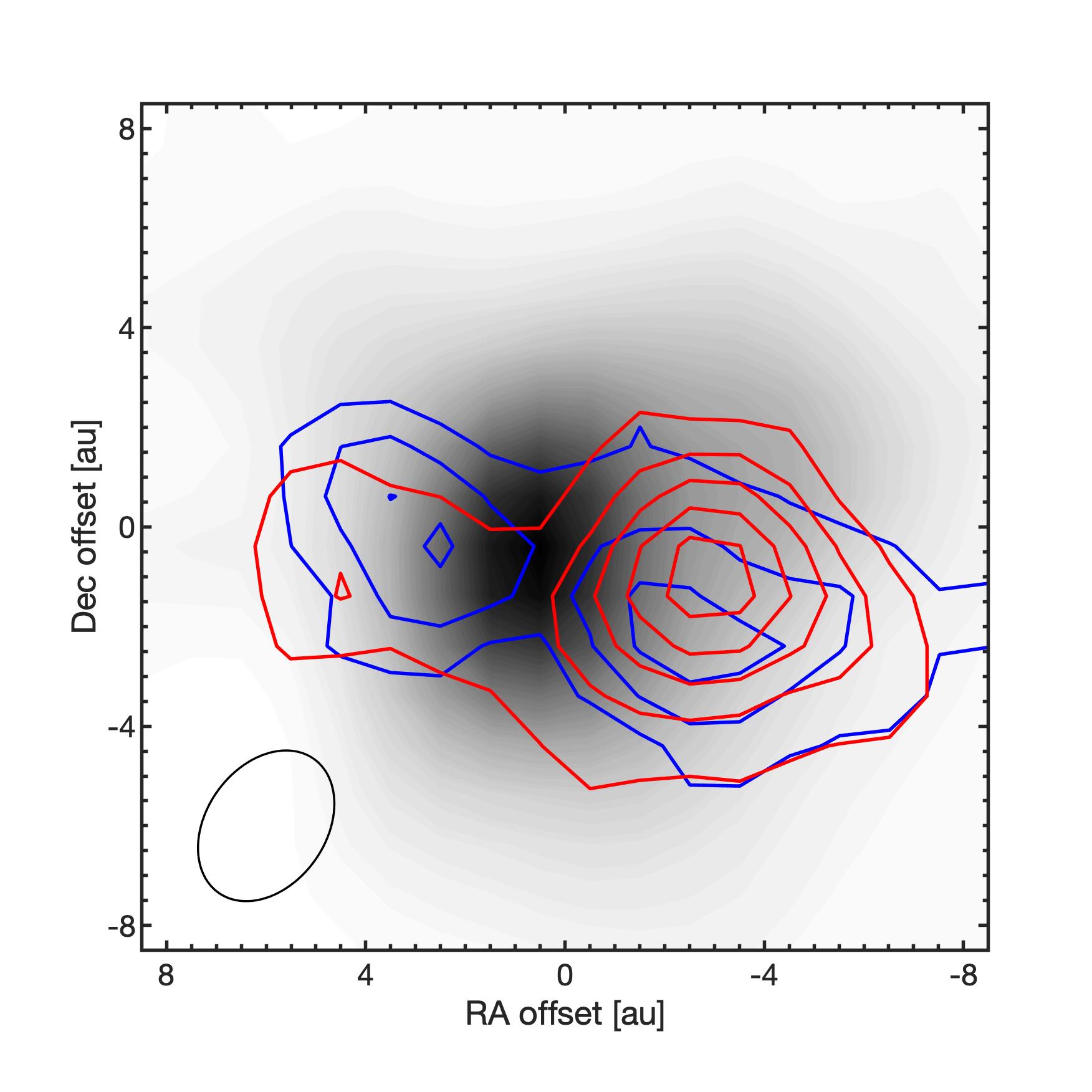}
   \caption{Integrated blue- and redshifted SiO (5--4) emission ($\pm 1 ~\rm{to}~ \pm 10$ \kmpers w.r.t.\ systemic velocity; contours) overlaid on the 1.3 mm continuum emission (greyscale). Contours are from 3$\sigma$ in steps of 1$\sigma$, where $\sigma$~=~2.1~mJy beam$^{-1}$\, \kmpers. The synthesized beam is shown in the lower left corner of the map.
   }
   \label{fig:siovscont}
\end{figure}

Along with the observation of the CO isotopologues, one of the SPWs in our long-baseline data set covered the frequency range 216.6 -- 217.6 GHz centred on the SiO\,(5--4) emission line. Fig.\ \ref{fig:siovscont} shows our detection of SiO, overlaid on the continuum emission. SiO was only recently detected in the vicinity of the B335 protostar \citep{Imai:2019ip}. Those observations were acquired at a spatial resolution of 0.1\arcsec, and no extension in the outflow direction could be resolved. \citet{Imai:2019ip} suggests that SiO could be tracing the launching point of the outflow or the accretion shock of infalling material. In the higher resolution observations of this paper, however, we spatially resolve the SiO emission and find that it is elongated along the outflow direction on both sides of the protostar. Since, theoretically, magnetically-powered jets are expected to rotate \citep[e.g.][]{Blandford:1982fj}, we searched for, but could not find, any signs of rotation in SiO along the expected jet axis. This is most likely because the SiO emission is not well-resolved in the north-south direction (the FWHM of the SiO emission in the direction perpendicular to the outflow axis is comparable to the size of the beam). 

Serendipitously, in the same SPW, we detected CH$_3$OH 5$_{-1,4}$--4$_{-2,3}$E (v$_t$=0) at 216.945~GHz ($E_{\rm{up}}=56~\rm{K}$), CH$_3$OH 6$_{1,5}$--7$_{2,5}$A (v$_t$=1)  at 217.299~GHz ($E_{\rm{up}}=374~\rm{K}$), and SO$_2$\,22(2,20)--22(1,21) at 216.643~GHz within a circular region of radius $\simeq$15~au centred on the protostar.

Although, the emission from all of the aforementioned species suffers absorption towards the protostellar position, they are otherwise well represented by Gaussian 2D fits. Moment 0 maps and position-velocity (PV) diagrams of these transitions are found in Figs.\ \ref{fig:CH3OH_PV} and \ref{fig:a1}.

\subsubsection{Deriving the excitation conditions from \texorpdfstring{CH$_3$OH}{CH3OH}}
\label{subsub:ch3oh}
\begin{figure}
   \flushleft
   \includegraphics[width=\hsize, clip=true, trim=0cm 0cm 0cm 6.5cm]{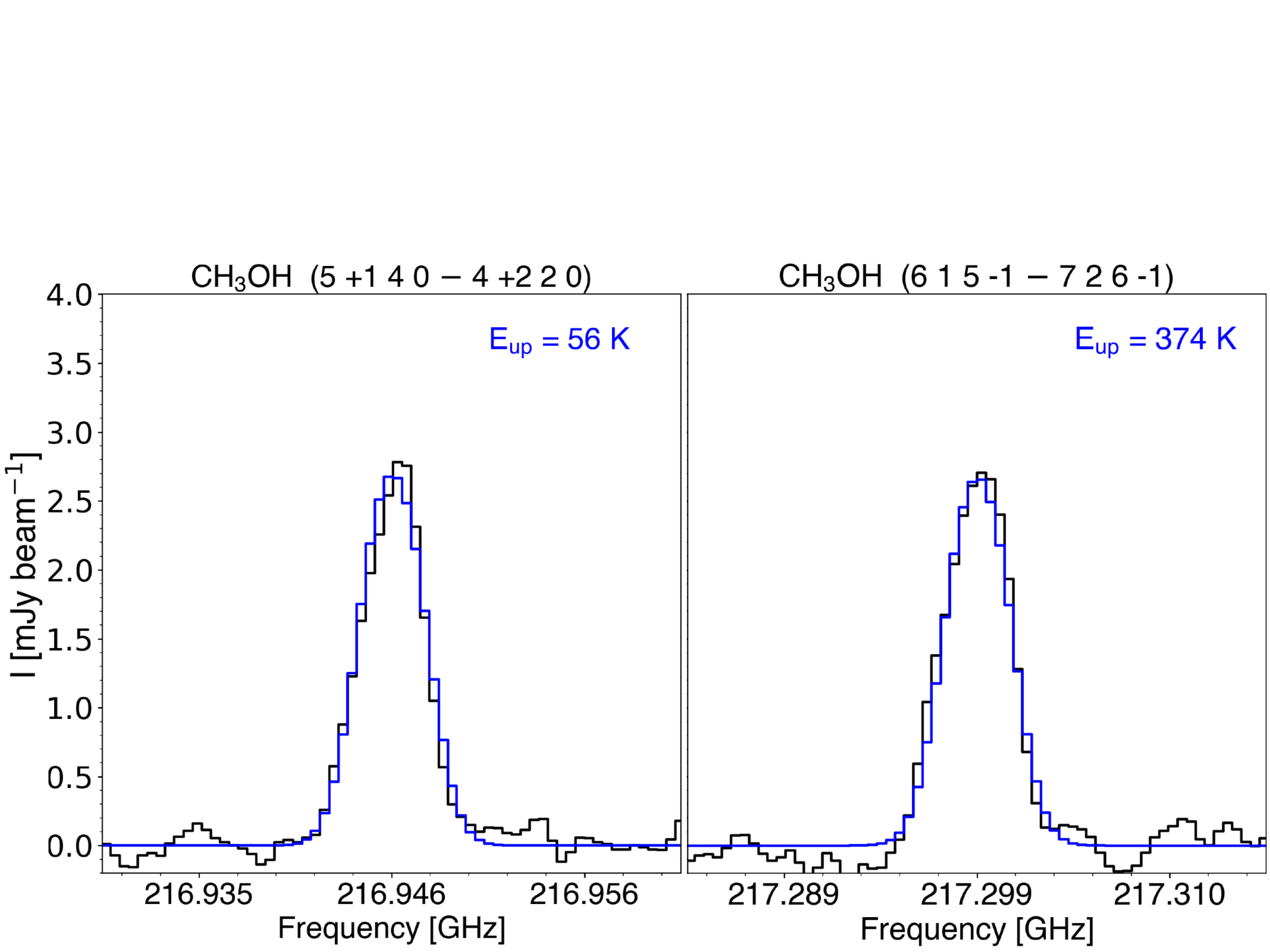}
   \caption{The CH$_3$OH 5$_{-1,4}$--4$_{-2,3}$E (v$_t$=0) and CH$_3$OH 6$_{1,5}$--7$_{2,5}$A (v$_t$=1) lines of CH$_3$OH detected towards B335 (black line) with an LTE spectral model overlaid in blue.}
   \label{fig:ch3oh}
\end{figure}

Four lines of CH$_3$OH fall within the frequency range of our long-baseline observations. Two of these lines, CH$_3$OH 5$_{-1,4}$--4$_{-2,3}$E (v$_t$=0) at 216.945~GHz and CH$_3$OH 6$_{1,5}$--7$_{2,5}$A (v$_t$=1) at 217.299~GHz are detected at a $\sim$20$\sigma$ level. The 6$_{-3,4}$--7$_{-1,6}$E (v$_t$=0) and 16$_{-1,15}$--15$_{-3,13}$E (v$_t$=0) lines are not detected. By combining the upper energy levels and line strengths of these four lines, we can strongly constrain the excitation temperature and column density of the CH$_3$OH emitting region with the spectral modelling code CASSIS\footnote{CASSIS has been developed by IRAP-UPS/CNRS: \url{http://cassis.irap.omp.eu/}}. Assuming local thermodynamic excitation (LTE) and optically thin emission, we ran a grid of models covering a range of column densities and excitation temperatures. The source size was taken to be 0.15\arcsec\ based on a 2D Gaussian fit to the CH$_3$OH emission (see Sect.\ \ref{subsub:infallandrotation}). The spectroscopic information was taken from the CDMS\footnote{The Cologne Database for Molecular Spectroscopy: \url{https://cdms.astro.uni-koeln.de/}} catalogue \citep{Muller:2001kr,Muller:2005al,Endres:2016ny} entry for CH$_3$OH. Figure \ref{fig:ch3oh} shows the observed spectra (in black) extracted from a 0.15\arcsec\ region centred on the emission peak of B335, as well as the best-fit spectral model (in blue). In order to reproduce the detected lines, and \emph{not} produce emission for the two non-detected lines, a CH$_3$OH column density of $6.8\pm0.1\times 10^{18}$\,cm$^{-2}$ and an excitation temperature of 220$\pm20$\,K is required. These lines have an optical depth, $\tau$, of $\sim$0.7. The CH$_3$OH emission shown in Fig.\ \ref{fig:CH3OH_PV} meanwhile indicates that the lines may be optically thick towards the protostellar position. However, the spectra extracted over the entire emitting region have line profiles with $\tau$~<~1, which suggests that the optically thin approximation is overall appropriate. The excitation temperature derived from the spectra is consistent with what is found on 5~au scales when using the previously estimated dust temperature of 30~K at 600~au \citep{Chandler:1993ms}, and assuming that its dependence on the distance from the protostar follows a power-law with index $-0.4$ \citep{Yen:2015vf,Shirley:2000qf}.
%

\section{Discussion}
\label{sec:discuss}
\subsection{The circum-protostellar environment}
\label{sub:circumstellarregion}

\subsubsection{Mass of the central emission}
\label{subsub:massofthecentralcomponent}
The circumstellar material traced by the continuum in our long-baseline observations is well-fitted by a 2D Gaussian with a FWHM of $\sim$7~au, and $S_{1.3 mm} = 28\pm0.2$ mJy. The continuum emission is therefore barely spatially resolved. The mass of the emitting region is estimated from
\begin{equation}
    M_{\rm{1.3 mm}} = \frac{S_{\rm{1.3 mm}}~d^2}{\kappa_{\rm{1.3 mm}} ~B(T_{\rm{dust}})},
    \label{eq:mdust}
\end{equation}
where $\kappa_{\rm{1.3 mm}}$ is the dust mass opacity at 1.3 mm, $d$ is the distance to B335, and $B(T_{\rm{dust}})$ is the Planck function at the temperature of the dust, $T_{\rm{dust}}$. To enable comparison with earlier results, we use a total mass opacity of $0.014~\rm{cm}^2 \rm{g}^{-1}$ \mbox{\citep{Yen:2015vf,Chandler:1993ms}}. This value is similar to tabulated values for MRN grains \citep{Mathis:1977lr} with thin ice mantles at high densities and a standard gas-to-dust mass ratio \citep{Ossenkopf:1994fk}, but is slightly lower than what is generally assumed for Class I and II disks \citep[e.g.][]{Ricci:2010ga,Harsono:2018lr}. Assuming that the gas temperature in the methanol emitting region is a good proxy for the dust temperature in the continuum emitting region (200~K; see Sec.~\ref{subsub:ch3oh}), the resulting circumstellar mass within 7~au is $M_{\rm{dust}}$~=~3\texpo{-4}~\msun. This mass is comparable to the 7.5\texpo{-4}~\msun\ inferred within a 25 au radius by \citet{Evans:2015qp}, and in the case of spherical symmetry, therefore, consistent with a power-law density distribution with index, $p = -2$ \citep{Yen:2015vf}. The circumstellar mass is two orders of magnitude smaller than the 0.05 -- 0.15~\msun \citep{Yen:2015vf,Evans:2015qp} that has accumulated in the central zone, as inferred from models of inside-out collapse \citep{Shu:1977lr}. Since the temperature gradients on even smaller scales cannot be measured with our current data, we note that the circumstellar mass could in fact be lower. For instance, by eq.\ (\ref{eq:mdust}), increasing the dust temperature by a factor of two yields a factor of two lower mass.

\subsubsection{Infall and rotation on small scales}
\label{subsub:infallandrotation}
\begin{figure*}
   \centering
   \includegraphics[width=0.66\hsize]{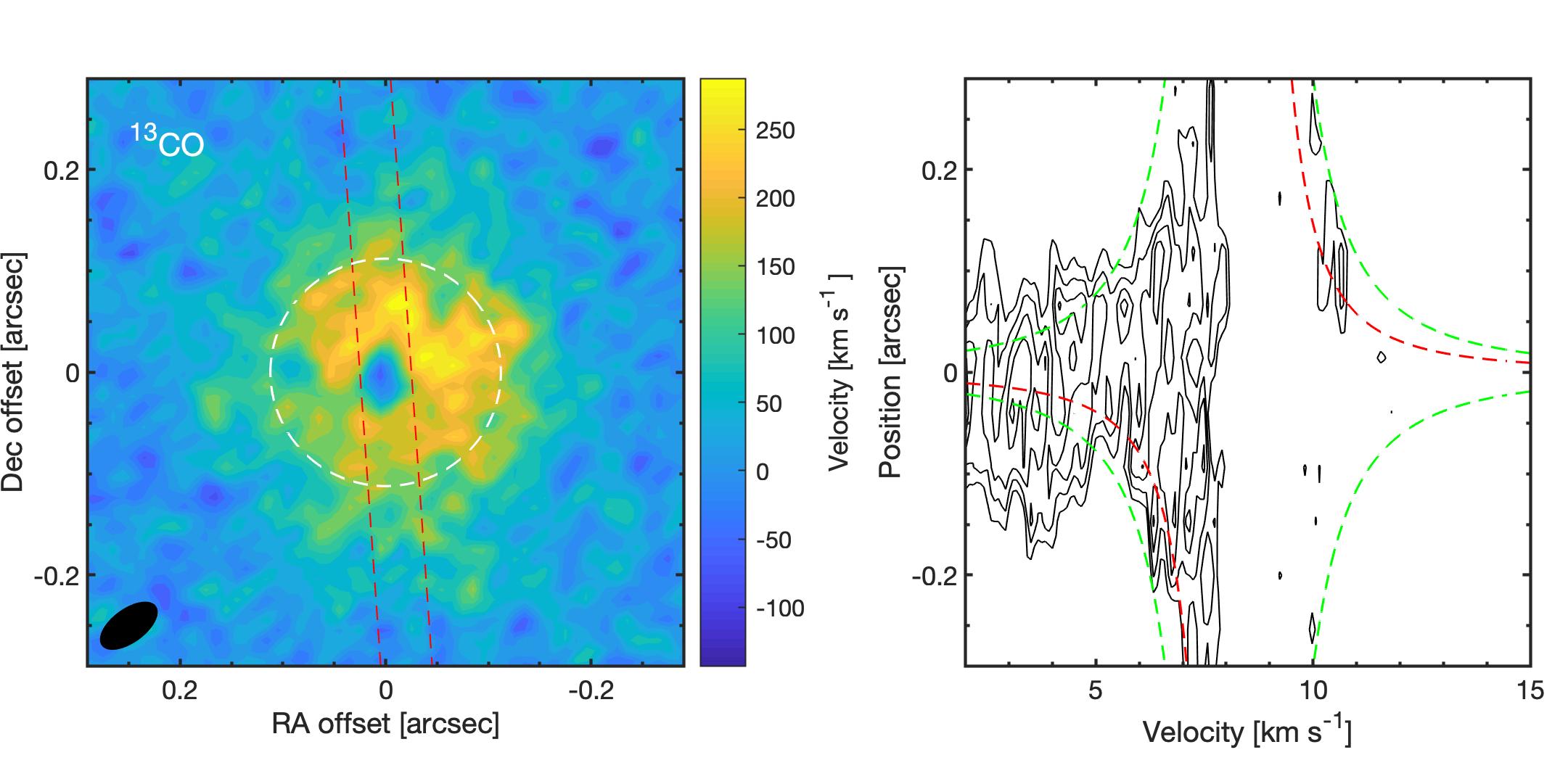} \\
   \vspace{-0.69cm} \includegraphics[width=0.66\hsize]{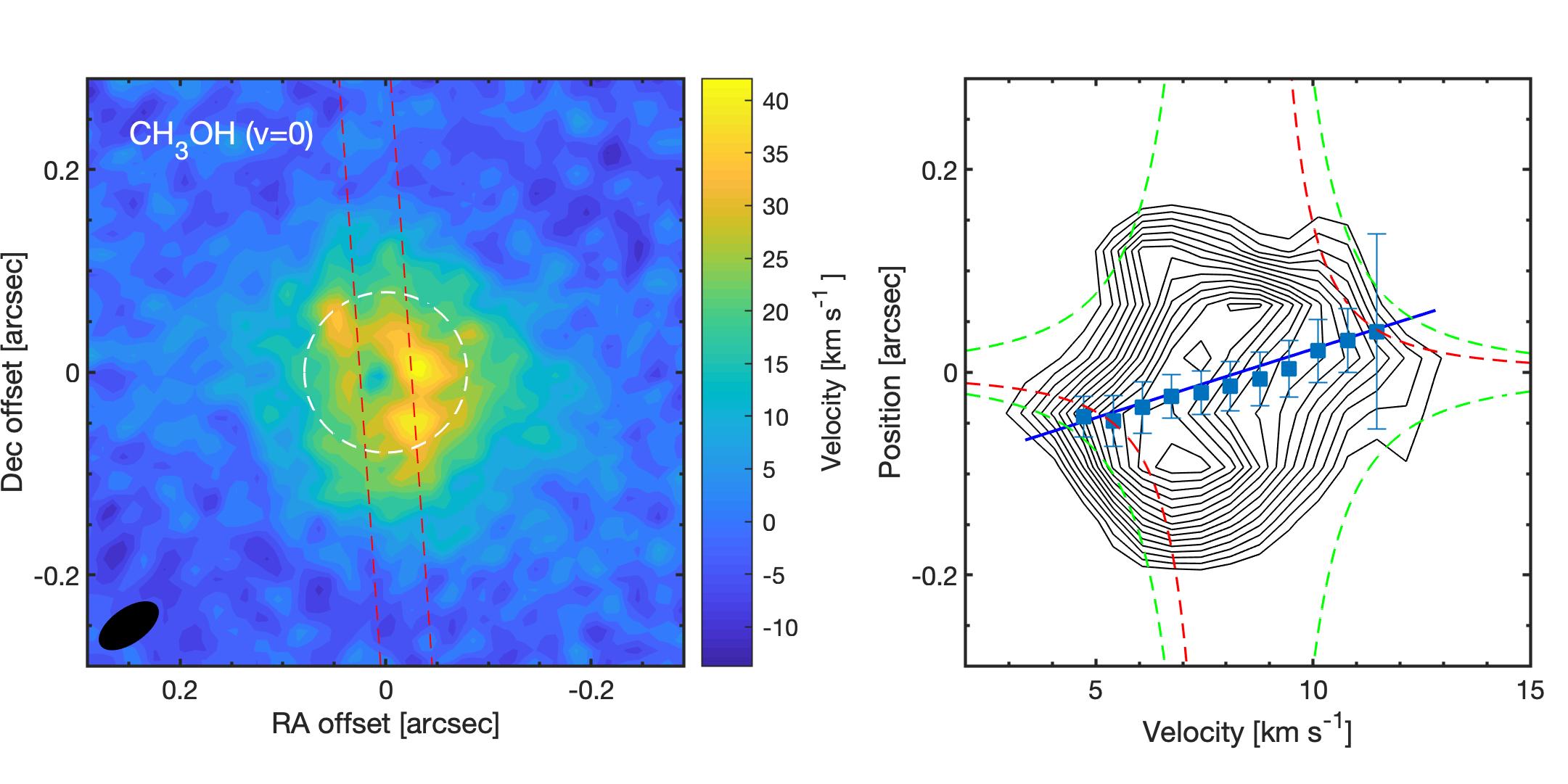} \\
   \vspace{-0.69cm} \includegraphics[width=0.66\hsize]{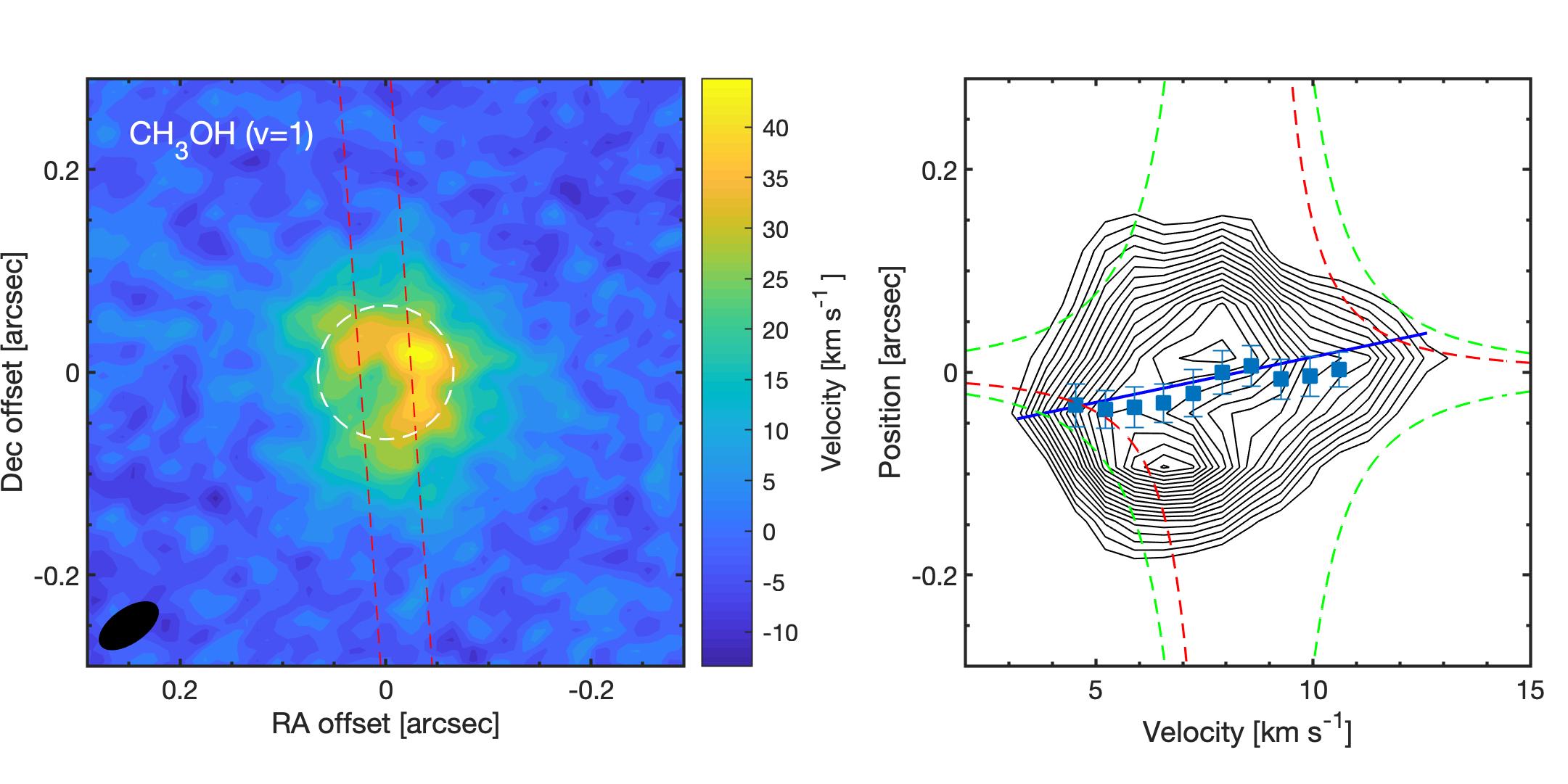} \\
   \vspace{-0.69cm} \includegraphics[width=0.66\hsize]{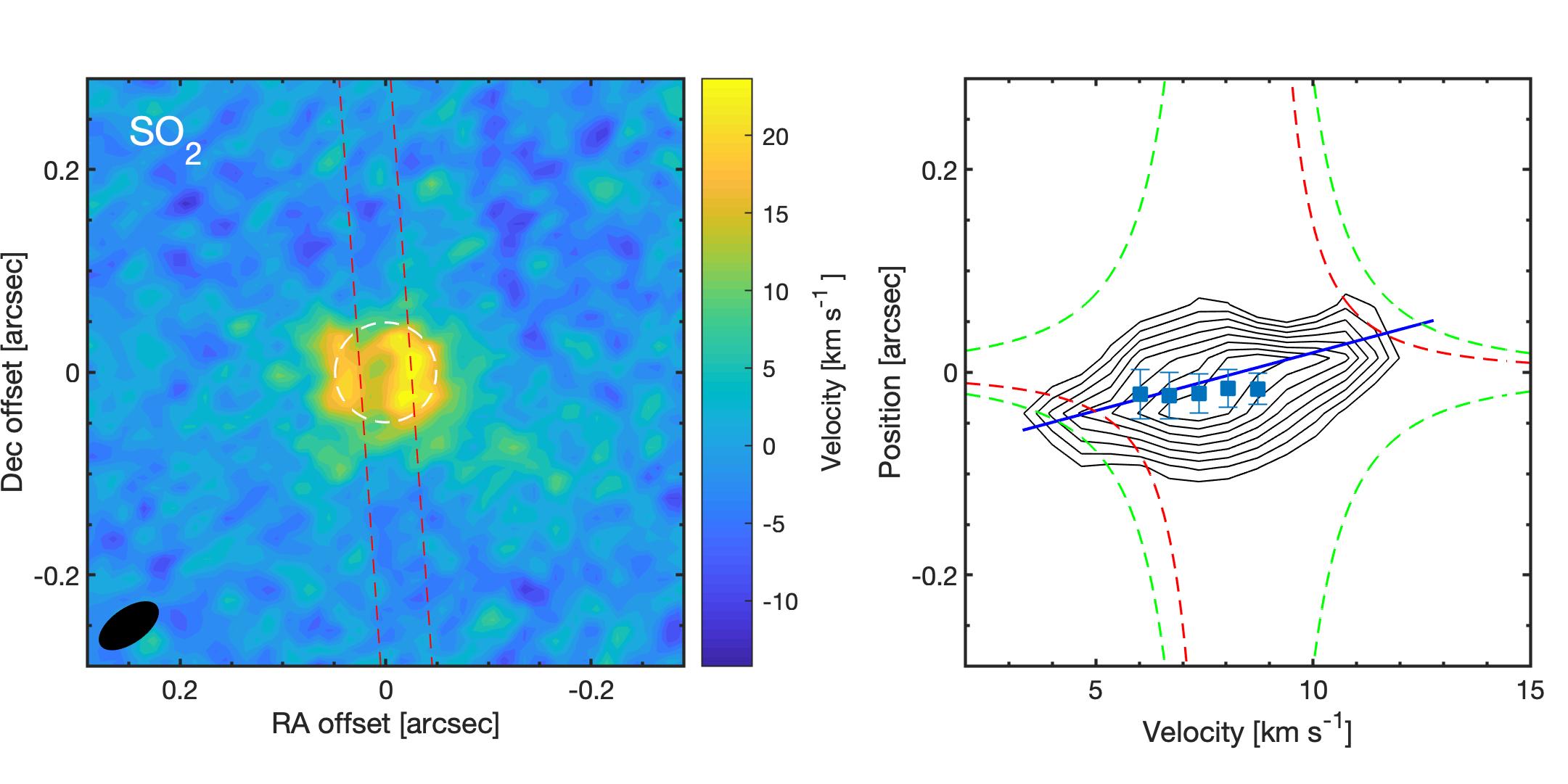} 
   \caption{\trettenco, CH$_3$OH (v=0,1), and SO$_2$ integrated emission and PV diagrams. \textit{Left} panels show emission integrated over 10~\kmpers\ centred on the systemic velocity. White dashed circles denote the FWHM of 2D Gaussian fits to the emission, while the red dashed lines indicate the direction and width of the PV cuts. The same PV cut is used in all panels. Synthesized beams are shown in the lower left corners. \textit{Right} panels show the resulting PV diagrams. Blue dots denote the location of emission peaks at a particular velocity (only velocity channels where the coefficient of determination, R$^2$, is larger than 0.3 are included). A linear fit to the peak emission velocity profile is represented by the blue line. Pure free-fall and pure Keplerian rotation velocities towards/around a 0.05~\msun\ protostar are indicated in the right panels by the green and red dashed lines, respectively. 
   } 
   \label{fig:CH3OH_PV}
\end{figure*}

Position-velocity cuts in the direction perpendicular to the outflow axis and across the protostellar position are presented together with the integrated emission ($\pm5\, \kmpers$, w.r.t.\ the systemic velocity) in Fig.\ \ref{fig:CH3OH_PV}. The FWHM of the emitting regions are estimated from 2D Gaussian fits to the emission\footnote{Since the errors on the FWHM are lower than 1~au in all cases, the errors are not reported here.}.

The PV diagram of $^{13}$CO shows a pronounced blue-red asymmetry; the blueshifted component is clearly visible while the redshifted components is nearly entirely absent. The observed blueshifted emission is consistent with infall towards a 0.05~\msun\ point-source (green dashed lines in Fig.\ \ref{fig:CH3OH_PV}), and which is substantially larger than the estimated mass of the circumstellar region (3\texpo{-4}\,\msun), but comparable to the mass estimated from inside-out collapse (0.05-0.15\,\msun; see Sect.\ \ref{subsub:massofthecentralcomponent}). A blueshifted, higher velocity component ($\lesssim$5~\kmpers) is visible in the PV diagram, and its offset from free-fall and Keplerian rotation curves suggest that it may be associated with the outflow. However, we find no significant signature of rotation in the \trettenco emission. The FWHM of the $^{13}$CO emitting region is estimated at 22~au, i.e., extended compared to the size of the beam.

The PV diagrams for the two CH$_3$OH transitions exhibit a distinctly different morphology and kinematics relative to \trettenco. The FWHM of the emitting CH$_3$OH regions are estimated at 16 and 13~au for v=0 and v=1 lines, respectively, which is significantly smaller than the region probed by \trettenco. The highest velocities observed in CH$_3$OH are found at small separations from the protostar, which is consistent with an infalling, rotating flow inferred from previous observations \citep{Yen:2015vf}. A rotation signature, consistent with solid-body rotation, is observed. However, the extent of the emission, plus the sensitivity limits on the higher velocity emission, do not allow us to significantly constrain the rotation profile. Given the current data, we can therefore not rule out Keplerian rotation around a 0.05~\msun\ protostar (red dashed lines in the right panels of Fig.\ \ref{fig:CH3OH_PV}). 

That said, we examined the velocity gradient in the immediate vicinity of the protostar in detail, and derived the peak position of the emission as a function of velocity by fitting Gaussians to the flux in each channel in the PV diagram. The ability of a fit to reproduce the flux profile in each channel is evaluated using the coefficient of determination (``R squared''), where a value of 1 indicates that the variance is fully accounted for by the fit. We present only velocity channels where R squared is larger than 0.3 and then subsequently fit a straight line through the peak positions (a lower threshold does not affect the results) assuming that the source velocity is 8.3~\kmpers\ w.r.t. \vlsr. The errors on the estimated peak emission positions are included as parameters in the fit. As shown in the right panels of Fig.\ \ref{fig:CH3OH_PV}, we arrive at velocity gradients that are perpendicular to the outflow axis with values of 1.4~$\pm$~0.3 and 0.9~$\pm$~0.5~\kmpers~au$^{-1}$ for the v=0 and v=1 transitions of CH$_3$OH, respectively. These values can be compared to the velocity gradient of 0.27~$\pm$~0.03~\kmpers~au$^{-1}$ derived from the \catteno emission on scales $\sim$30 au, as reported by \citet{Yen:2015vf}. These authors observed B335 at an order of magnitude coarser angular resolution than the current long-baseline observations. They are thus probing scales where any Keplerian rotation would be a factor of three smaller, while in the case of solid-body rotation, velocities would be a factor of ten higher. By combining the extent of the observed CH$_3$OH emission and the derived velocity gradient, we estimate the specific angular momentum within $\sim$20 au to be $\sim$25~au~\kmpers.

Due to the compact and clearly rotating nature of the methanol emission, and the fact that it rotates at a velocity that is a factor of three to four times higher than what was previously derived from C$^{18}$O at larger scales, we suggest that the methanol emitting region is associated with an inner, rotating, and possibly disk-like structure.

In the case of SO$_2$, using Gaussian fitting, we find the emission has a FWHM of $\simeq$10~au, i.e., even smaller than that of CH$_3$OH. We note that only a small amount of blueshifted emission is detected on the northern side of the source, and vice versa on the southern side; if the emission was predominantly tracing infall, we would expect to see blueshifted emission on the northern side of the PV diagram (and vice versa for the southern side). We believe the reason for this is that the rotational velocity component is dominant over the infalling velocity component in this region. It has recently been suggested that SO$_2$ can, in fact, be a tracer of accretion shocks close to the protostar \citep{Artur-de-la-Villarmois:2018bg}. Unfortunately, although SO$_2$ shows the most compact morphology of the species detected in these observations, we cannot, at present, draw any firm conclusions regarding its origin. What is clear, however, is that the observed gas is rotating on the very smallest scales.

\subsection{The outflow}
\label{sec:theoutflow}
\subsubsection{The cavity walls}
\label{subsub:extendeddust}
The combined observations presented here recover most of the \tolvco\ emission in the line wings relative to single-dish observations (cf.\ Fig.\ \ref{fig:apexcomp}). Thus, we claim that the majority of the \tolvco emission in B335 originates in the narrow arms of emission that follow the X-shaped cavity walls (Figs.\ \ref{fig:lbvs288} \& \ref{fig:moment0_combined}). Although the width of these arms (up to $\sim$50 au) vary slightly along the cavity walls, the width is almost always larger than our beam size, and most of the cavity emission is resolved. 

While the peak of the continuum emission is concentrated at the protostellar position in both long-baseline and combined data, only the combined data set reveals extended 1.3 mm emission (Figs.\ \ref{fig:moment0_combined} and \ref{fig:a0}) to the north and south. The most striking feature of the continuum (greyscale; Fig.\ \ref{fig:a0}) is that it stretches $\sim$800 au to the north (i.e.\ greater than the extent of Fig.~\ref{fig:moment0_combined}) and a few hundred au to the south. In addition, the map shows that the dust emission follows the cavity walls, especially to the south-west, in agreement with \citet{Maury:2018qf}. However, it is only prominent at separations from the outflow axis which are greater than for the \tolvco emission, and the apparent opening angle (when examining the brightest dust continuum emitting regions to the east) is slightly larger than the opening angle of the \tolvco\ emission. Although we cannot entirely refute that the dust is being carried by the outflow, the coincidence with the X-shaped morphology of the \tolvco, suggests that the wind is excavating a cavity and/or entraining dust from the envelope, rather than transporting it from the circum-protostellar region. Under the reasonable assumption that a significant fraction of the \tolvco\ emission in the cavity walls (which extends for $\gtrsim$2000 au) is caused by mechanical interaction between the outflow and the envelope, the cavity wall emission provides an estimate on the lateral scales over which entrainment takes place. 

That said, while the edge of the continuum emission is in close proximity of the \tolvco\ emission, it is possible that some of the emission observed in the north and south is actually tracing the infalling surrounding envelope. In any case, to our knowledge, B335 is the only protostellar example where dust continuum is detected towards an outflow cavity wall.

\subsubsection{Outflow rotation and cavity expansion}
\label{subsub:outflowrotation}
Since Keplerian rotation has not been detected towards B335 on scales larger than 10~au, it is interesting to instead search the long-baseline data for signs of outflow rotation. Unfortunately, the \tolvco line profiles are dominated by infall and outflow motions and we find no significant evidence of rotation in the outflow. As mentioned in Sect.\ \ref{sub:resultsSiOCH3OHSO2}, however, CH$_3$OH and SO$_2$ do show clear evidence for rotation within $\sim$16~au (Fig.\ \ref{fig:CH3OH_PV}). This emission originates in the central region, where it appears that material is infalling and rotating simultaneously. Even so, the velocity gradients measured from the CH$_3$OH and SO$_2$ lines are consistent with one another, and thus one might expect that the outflow should rotate in the same sense.

A \tolvco moment 1 map (Fig.~\ref{fig:a2}) and line profiles at different positions (Fig.\ \ref{fig:moment0_combined}), meanwhile, do not show any significant difference in the velocities between northern and southern outflow cavity walls. In the redshifted outflow, however, there is a blueshifted component at small separations ($\lesssim$30~au) south-west of the protostar (Fig.\ \ref{fig:lbvs288}), consistent with the inferred rotation seen in SO$_2$ and CH$_3$OH. Likewise, a redshifted component is observed in the blueshifted outflow on comparable scales, but this component is seen along the jet axis and not the north-western side as would be expected if this component was due to rotation in the outflow.

\begin{figure*}[ht!]
   \centering
   \includegraphics[width=0.75\hsize]{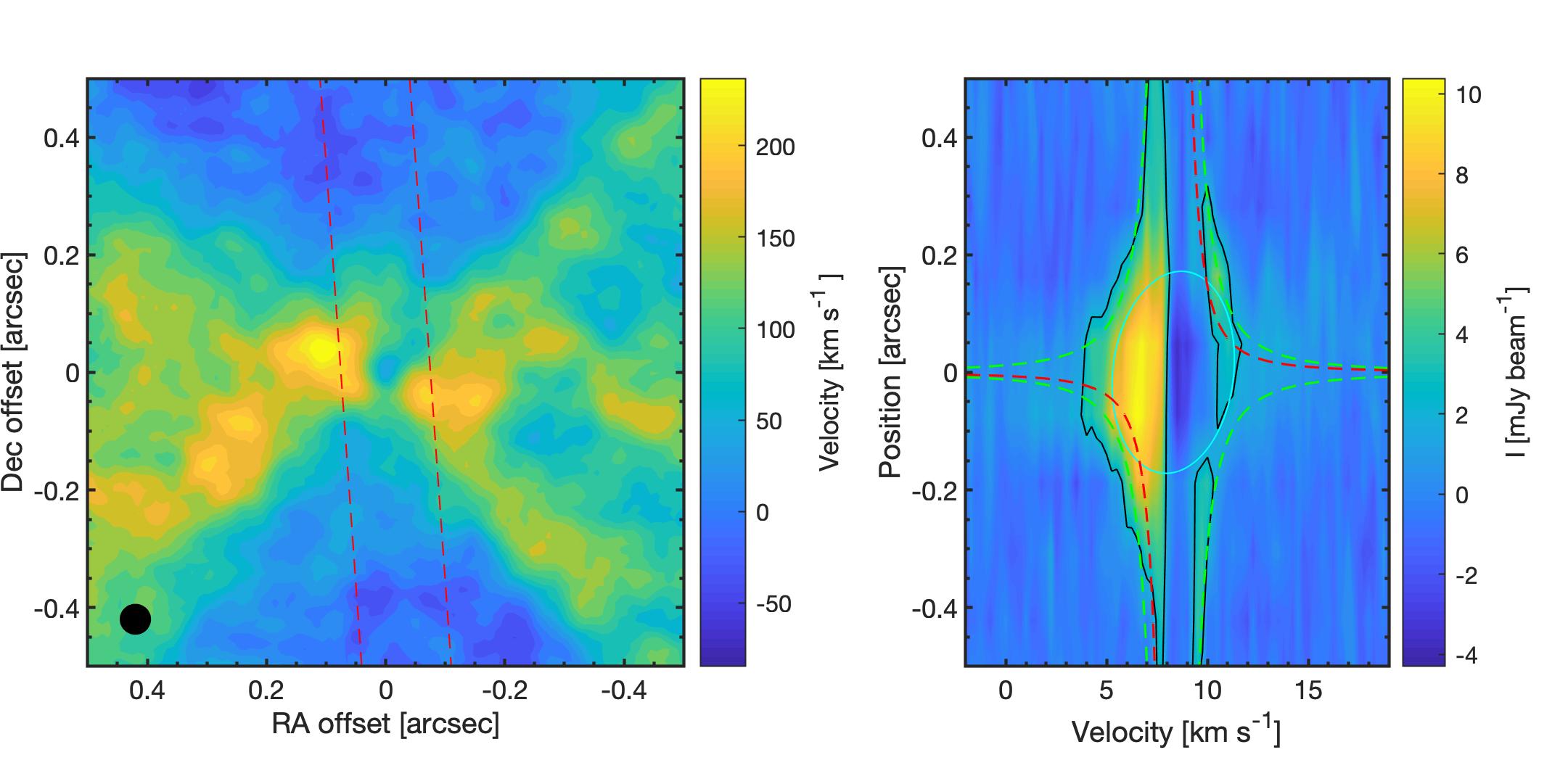} \\
    \includegraphics[width=0.75\hsize]{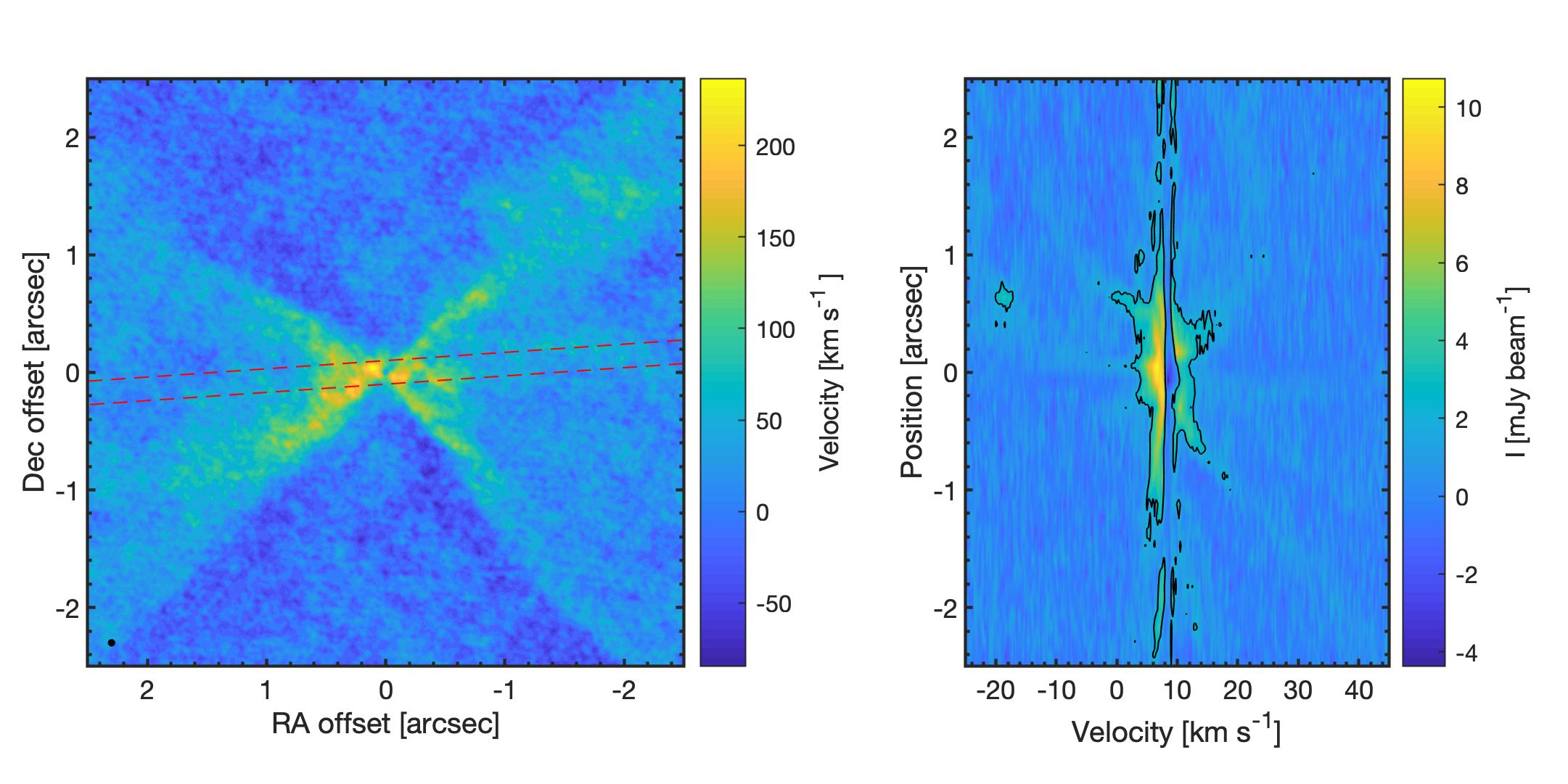} \\
    \includegraphics[width=0.75\hsize]{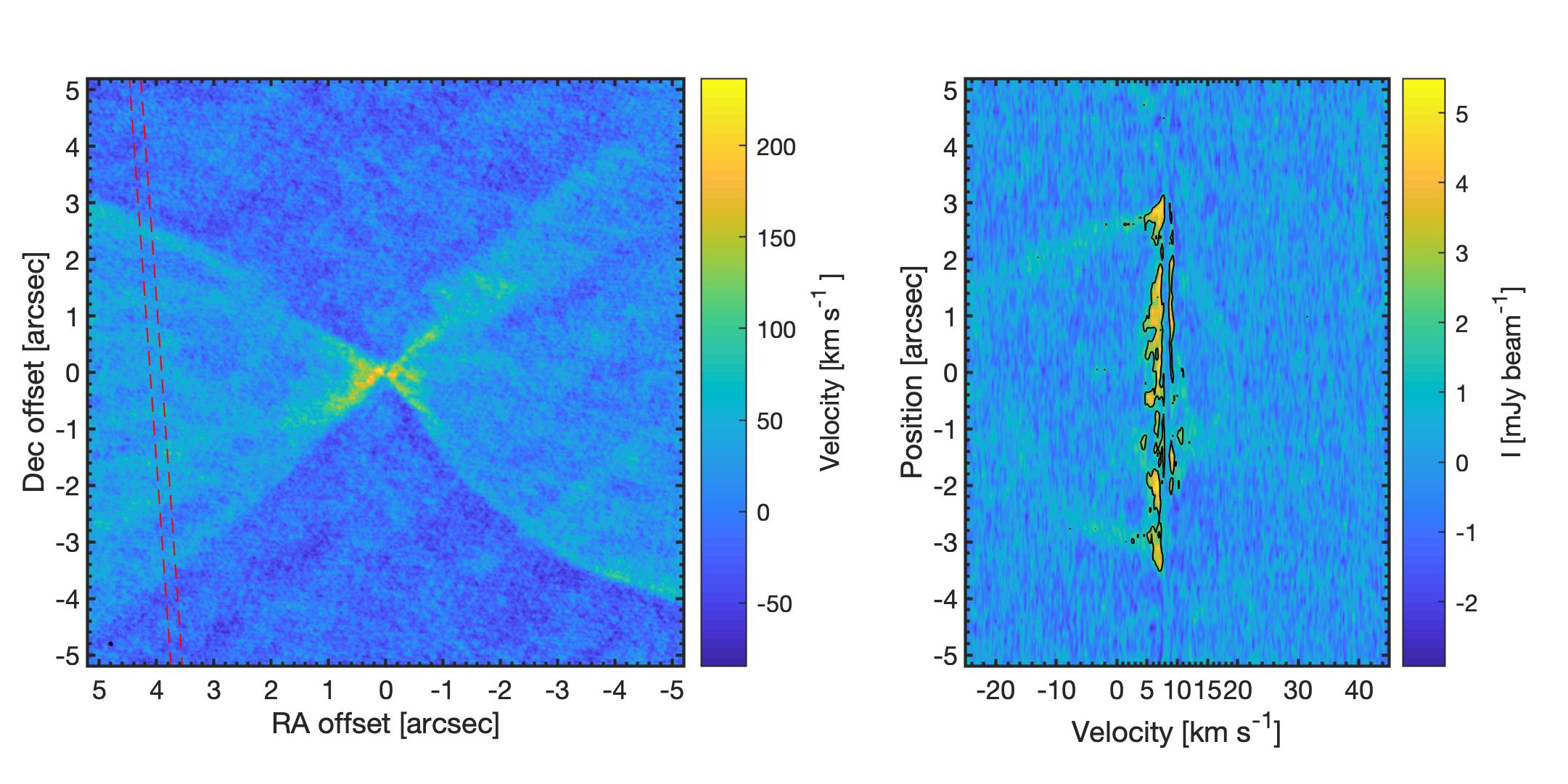}
    \caption{\tolvco integrated emission and PV diagrams. \textit{Left panels:} Moment 0 maps of the \tolvco emission (within~$\pm~5~\kmpers$ from the systemic velocity) overlaid with the orientation and width of each PV cut. The maps were convolved with a 2D Gaussian to 5~au resolution to improve the S/N ratio. \textit{Right panels:} PV diagrams for \tolvco along the different cuts shown in the left panels. Contours are at 3$\sigma$. In the upper right panel, green and red dashed lines denote pure free-fall and Keplerian rotational velocities towards/around a $0.05\msun$ protostar, respectively. An example of a 2D Gaussian fit (see Sect.\ \ref{subsub:outflowrotation}) to the emission is indicated by the cyan ellipse. Note the change in scale from top to bottom rows.}
    \label{fig:12COPV}
\end{figure*}

Attributing the inner features to rotation is in any case difficult since the outflow is nearly in the plane of the sky, and expansion of the outflow, deviations of the outflow axis from axisymmetry, or internal shocks \citep[e.g.][]{Fendt:2011tk} can produce similar features. Hints of expansion are indeed seen in the lowermost PV diagram of Fig.\ \ref{fig:12COPV}, where an extended U-shaped structure is observed at blueshifted velocities. Although this emission structure barely exceeds the 3$\sigma$ threshold, it is clear that the velocity is increasing towards the outflow axis. A similar feature is found on the redshifted side at slightly lower S/N, and lower velocities. While an onion-layered velocity structure (whereby velocities close to the outflow axis are high and then fall off with distance) is expected for a rotating, magnetically-launched wind \citep[e.g.][]{Bacciotti:2002fr,Pudritz:2006kq}, the symmetry of the PV diagram w.r.t.\ the zero position offset in the lowermost panel of Fig.\ \ref{fig:12COPV}  is not consistent with this picture. It is, however, consistent with expanding cavity walls.

The rotation and expansion velocity at different offsets from the protostar can be estimated using the method presented in \citet{Lee:2018et}, assuming that only two velocity components are present. While this is an oversimplification of reality, assuming the outflow velocity does not change significantly over the considered distances, it does provide a useful method to deduce any trends in rotation and expansion. 

We fit 2D Gaussians to several PV diagrams with cuts perpendicular to the outflow axis but with different offsets. One example of such a fit (at zero offset from the continuum peak) is presented in the uppermost panel of Fig.\ \ref{fig:12COPV}). We only consider pixels where the emission is above 3$\sigma$, and we purposefully mask out velocities between 7.3 and 9.3~\kmpers, i.e., where emission from the foreground is expected to be significant\footnote{We note, however, that including velocities from 7.3 -- 9.3~\kmpers in our analysis does not affect our results.}. With this approach, the rotation velocity is given by the 1$\sigma$ velocity at maximum extension with respect to the central velocity of a Gaussian fit to the emission. Meanwhile, the expansion velocity is given by the difference between the velocity at the zero position offset and the central velocity of the Gaussian fit.

\begin{figure}
   \flushleft
   \includegraphics[width=1.0\hsize]{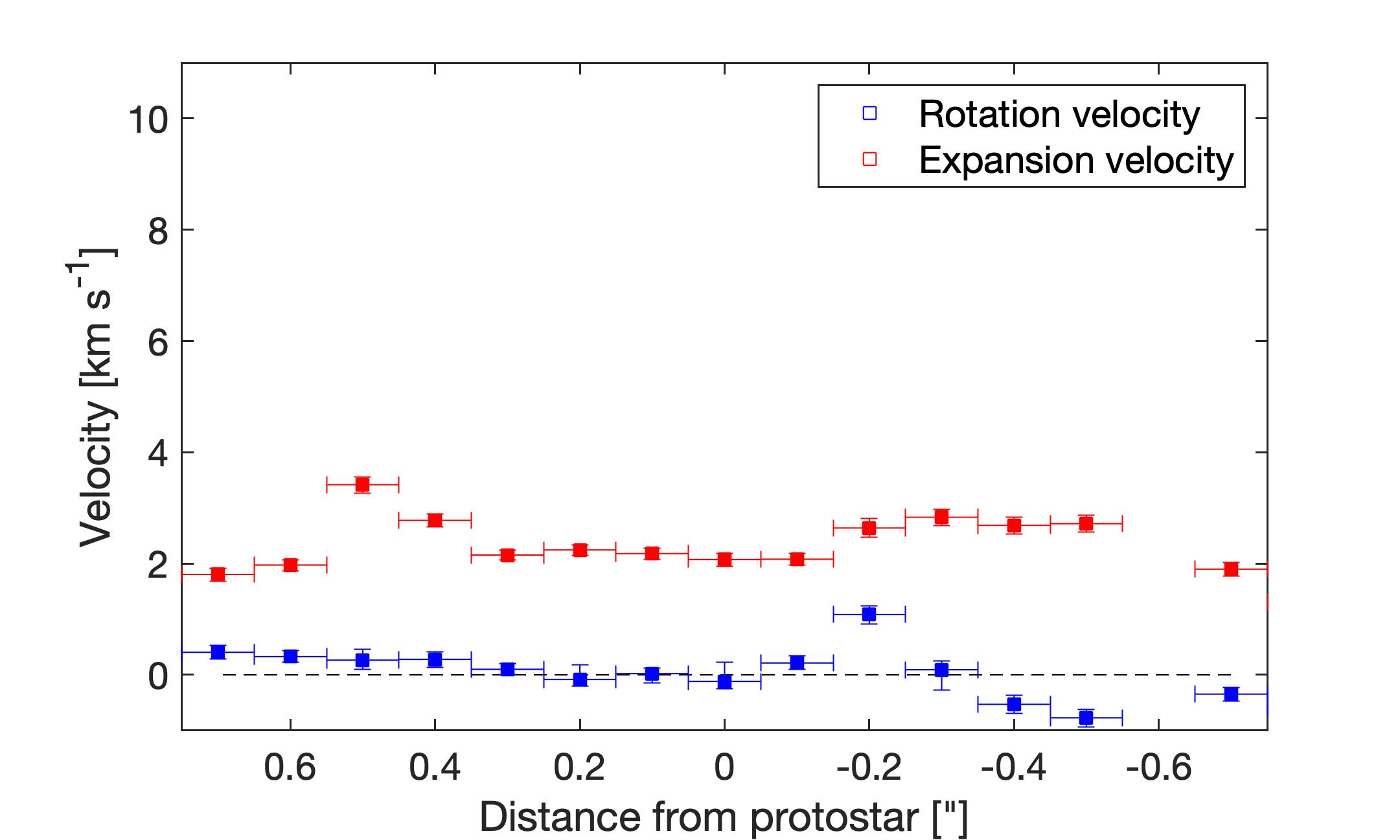}
   \caption{Rotation and expansion velocities of the outflow derived from PV cuts of the \tolvco emission across the outflow axis with different separations from the protostar. Error bars are given by the uncertainty in the 2D Gaussian fits and the width of the PV cut.}
   \label{fig:rotexpvel}
\end{figure}

Figure \ref{fig:rotexpvel} shows the results of the fitting procedure, but clear evidence for rotation in the outflow cannot be inferred from this analysis except at $\lesssim$20~au separation from the protostar in the redshifted outflow, where the values are on the 1~\kmpers level. Since any rotational motions of the \tolvco outflow are apparently at a minor level, and given that outflows, if magneto-centrifugally-powered, are expected to rotate \citep{Blandford:1982fj,Pudritz:1983fv}, it is reasonable to assume that a large fraction of the CO emitting gas is instead material that is being entrained by the outflow.

In addition, the inferred expansion velocities in the outflow, $\sim$(2~--~4~\kmpers), are at least an order of magnitude lower than what is needed to explain the observed width of the outflow given the forward velocities previously inferred from proper motion of the HH objects, assuming its width is due entirely to expansion.

Considering these results, we suggest that what we predominantly are seeing in \tolvco is the excavation of a cavity and entrainment of material, rather than the wind itself.

\subsubsection{Launching of the wind}
\label{subsub:windlaunching}
The idea that protostellar jets are launched magneto-centrifugally from disks is now widely accepted, although several unresolved issues remain (see, e.g. \citealt{Frank:2014ys}). In our long-baseline data, the intersection point of the X-shaped morphology of the \tolvco emission is consistent with small outflow launching radii. This is supported by the scale of the detected SiO emission that we associate with a central jet. The south-western and north-eastern cavity walls exhibit little curvature, while the other two cavity walls show no evidence for curvature at all. In the case of a magnetically-launched outflow, not only is the same mechanism responsible for the transfer of both kinetic energy and angular momentum into the wind, the outflow rotational and forward velocity components should also be closely linked \citep[e.g.][]{Ferreira:2006bh, Ramsey:2019ck}. This remains true whether the underlying mechanism is a disk wind and/or an X-wind \citep{Shu:1994kx}.

The middle panels of Fig.\ \ref{fig:12COPV} presents the PV diagram along the outflow for the combined \tolvco\ data set, and reveals some interesting features. First, at small positional offsets (within 100~au from the protostar), a single, significant, isolated high velocity blueshifted component can be seen at velocities reaching $\sim$30~\kmpers w.r.t.\ the systemic velocity. In addition, both blue- and redshifted components show marginally detected emission which extends out to $\sim$20~\kmpers and offsets of $\sim$100 au. These velocity components are likely associated with a small scale jet that is expected along the central axis of the outflow. Second, in the redshifted outflow, a feature is visible at offsets of up to 50 au, corresponding to a marginally detected high velocity feature visible in Fig.\ \ref{fig:moment0_combined} (red point/spectrum).

Although we do not detect any rotation in the outflow itself, the detection of rotation in CH$_3$OH on the smallest scales allows us to put an upper limit on the rotational velocity component of the order of 1~\kmpers. One can thus estimate the launching radius if the protostellar mass is known. Following \citet{Anderson:2003jk}, we use a velocity of 1.0~\kmpers and a maximum line-of-sight velocity of $\simeq$30~\kmpers, estimated from the aforementioned structures seen in the \tolvco PV cuts taken along the outflow at $\sim$60~au separation from the central source (where the velocities are higher then the local escape velocity). The width of the outflow in this region is of the order 80 au. We adopt an inclination angle 10$^\circ$ \citep{Hirano:1988lr} with respect to the plane of the sky, yielding a maximum \tolvco velocity of $\sim$170~\kmpers in the outflow direction. Applying Eq.\ (4) of \citet{Anderson:2003jk}, we thus estimate an upper limit for the launching radius of the high-velocity CO gas of the order of 0.07~au. Decreasing the rotation velocity by a factor of three implies a factor of two smaller launching radius ($r_0 \propto \upsilon_{\phi}^{2/3}$). We also note that a smaller inclination angle w.r.t.\ the plane of the sky implies even smaller launching radii.

The lack of a clear rotation signature in the \tolvco outflow prevents us from calculating the specific angular momentum in the regions of high velocity. We can, however, provide an upper limit of 40 au \kmpers\ (maximum rotational velocity is estimated at 1~\kmpers within 30~au from the protostar, where the width of the outflow is less than 40~au), which is consistent with the angular momentum derived from the observed rotation in methanol, viz.\ 25~au~\kmpers (Sect.\ \ref{subsub:infallandrotation}).

\subsubsection{Recent ejection outburst?}
\label{subsub:recentejection}
One way to characterise the accretion history of protostars is to study how the local chemistry is affected by a short burst in accretion luminosity \citep[e.g.][]{Bjerkeli:2016wq}. However, a disadvantage of this method is that it is only sensitive to relatively long time scales \citep[e.g.\ 100 -- 1000 yr;][]{Jorgensen:2013lr}. An alternative approach is to take advantage of the fact that increased accretion should also lead to enhanced outflow activity \citep[e.g.][]{Raga:1990ij}. This method has previously been used on large spatial scales \citep[e.g.][]{Plunkett:2015xy} to constrain the accretion history of protostars. With our current data, we can now apply the same analysis to very small spatial scales, and hence, on small temporal scales.

Emission from \tolvco is clearly detected at $-30$~\kmpers\ w.r.t.\ systemic velocity ($\sim$170\kmpers assuming an inclination of 10$^\circ$) and $\sim$60~au separation in the blueshifted outflow (Fig.\ \ref{fig:12COPV}; middle). The cause of this emission is not readily apparent from our data. It could be the continous, molecular component of a jet which is propagating along the outflow axis. This is not likely, however, since it is then not easy to explain why the emission is so spatially confined, and why it is only detected at an offset of $\sim$60~au.  Alternatively, it could be due to a wobbling/precessing jet and there is therefore a significant component that is not confined to the plane of the sky. However, we find this scenario unlikely as well because of the overall X-shaped morphology of the outflow and the fact that the high velocity feature is located only at a distance ($\sim$60~au) from the protostar. Instead, the explanation we find most plausible is that it is a molecular ``bullet'' associated with a recent accretion event and subsequent transient increase in ejection activity.

We interpret the features visible in the middle panels of Fig.\ \ref{fig:12COPV} as evidence for such episodic ejection. For the feature at $-30$~\kmpers\ w.r.t.\ systemic velocity and $\sim$60~au separation in the blueshifted outflow, we can estimate its dynamical timescale: Deprojecting with cos(\textit{i}) (assuming \textit{i}~=~10\adeg; \citet{Hirano:1988lr}), the speed and distance of this feature from the continuum peak implies that ejection took place within the last few years (formally, 1.7 years). This result is conspicuous. If outflow ejection was significantly enhanced only 2 years ago, it could be so that B335 very recently underwent a burst in accretion as well. This motivates a search of not only archival data, but also follow-up studies in the near future to monitor the accretion via episodic ejection events. It should be noted in this context, however, that if the inclination is closer to the plane of the sky \citep[e.g. 3\adeg;][]{Stutz:2008fk}, that would imply even shorter dynamical time-scales.

\section{Summary \& Conclusions}
\label{sec:conclude}
The isolated protostar B335 was observed using ALMA in its largest configuration ($\sim$16 km baselines). Not only are these the highest resolution observations of B335 to-date ($\sim$3~au resolution), by combining the long-baseline data with publicly available archival data, we are also able to probe large and small scales with high-fidelity simultaneously.

Dust continuum emission at 1.3 mm is seen in the long-baseline data towards the central position with a FWHM = 7~au, slightly larger than the size of the synthesized beam of the observations. The dust mass of the central emission component is estimated to be $M_{\rm{dust}}=$3\texpo{-4}~\msun, consistent with earlier results.

Lines of CH$_3$OH and SO$_2$ are detected in the vicinity of the protostar (FWHM $<$15 au) and show a rotational velocity gradient on the 1~\kmpers\ au$^{-1}$ level, corresponding to a specific angular momentum of $\sim$25~au~\kmpers. The PV diagrams for CH$_3$OH and SO$_2$ are consistent with Keplerian rotation around a protostellar mass of 0.05~\msun. From LTE analysis of the CH$_3$OH emitting region, we estimate a gas temperature of 220$\pm$20~K and a column density of 6.8$\pm$0.1\texpo{18}~cm$^{-2}$.

\tolvco\ emission from the outflow shows an X-shaped morphology where most of the emission is detected in narrow arms along the cavity walls (Fig. \ref{fig:moment0_combined}). Comparison with single-dish observations suggests that the majority of the CO emitting gas is in the arms and is being entrained over a $\sim$50~au thick region. 1.3 mm dust emission is also detected in the proximity of the presumed outflow cavity walls. Whether this dust is associated with the wind or excavated by the wind is, at present, not entirely clear. The coincidence between the continuum emission and the X-shaped morphology of the \tolvco emission, combined with a lack of evidence for rotation along the ``X'', implies the dust is locally being excavated and/or entrained rather than having originated close to the protostar and transported in the outflow. Furthermore, the \tolvco\ emission does not reveal any obvious signs of rotation except on the very smallest scales where a blueshifted component is detected in the redshifted outflow to the south. Position-velocity cuts in the direction perpendicular to the outflow further suggest that, if the outflow is rotating, it is at a very low level. We also estimated the values of the expansion velocity (Fig.\ \ref{fig:rotexpvel}), but the values are not sufficient to explain the observed width of the outflow.

Within $\sim$60~au of the protostar, episodic features moving at high velocities ($\sim$30~\kmpers) w.r.t.\ the systemic velocity are detected. The dynamical timescale of these knots is less than a few years. We estimate the launching radius of these episodic structures to be smaller than 0.1~au. In addition, SiO emission is observed to be elongated along the direction of the outflow and is likely associated with a jet, but does not show any clear signs of rotation on the scales probed here.

Taking all of the evidence together, and in particular the absence of a clear Keplerian disk on the smallest scales, suggests that B335 is either very young and/or disk growth is actively being inhibited by magnetic braking.

\begin{acknowledgements}
      We would like to thank Neal Evans for a throrough referee report that greatly helped improve the quality of this paper.
      
      This paper makes use of the following ALMA projects data: 2013.1.00879.S, 2016.1.01552.S, and 2017.1.00288.S. ALMA is a partnership of ESO (representing its member states), NSF (USA) and NINS (Japan), together with NRC (Canada), MOST and ASIAA (Taiwan), and KASI (Republic of Korea), in cooperation with the Republic of Chile. The Joint ALMA Observatory is operated by ESO, AUI/NRAO and NAOJ. \\

      We acknowledge support from the Nordic ALMA Regional Centre (ARC) node based at Onsala Space Observatory. This paper also makes use of data acquired with the Atacama Pathfinder EXperiment (APEX) telescope. APEX is a collaboration between the Max Planck Institute for Radio Astronomy, the European Southern Observatory, and the Onsala Space Observatory. Swedish observations on APEX and the Nordic ARC node are supported through Swedish Research Council grant No.\ 2017-00648.\\

      PB acknowledges the support of the Swedish Research Council (VR) through contracts 2013-00472 and 2017-04924. The computations were performed on resources at Chalmers Centre for Computational Science and Engineering (C3SE), provided by the Swedish National Infrastructure for Computing (SNIC). Part of this work was supported by the German \emph{Deut\-sche For\-schungs\-ge\-mein\-schaft, DFG\/} project number Ts~17/2--1. JPR was supported, in part, by the Virginia Initiative on Cosmic Origins (VICO) and, in part, by the National Science Foundation (NSF) under grant nos.\ AST-1910106 and AST-1910675. The research of LEK is supported by a research grant (19127) from VILLUM FONDEN. JKJ is supported by the European Research Council (ERC) under the European Union's Horizon 2020 research and innovation programme through ERC Consolidator Grant ``S4F'' (grant agreement No~646908). Research at Centre for Star and Planet Formation is supported by the Danish National Research Foundation (DNRF97).
\end{acknowledgements}

\bibliographystyle{aa}
\bibliography{papers.bib}
\clearpage
\begin{appendix}
\section{Supplementary material}
\begin{figure*}[ht!]
   \centering
   \includegraphics[width=0.55\hsize]{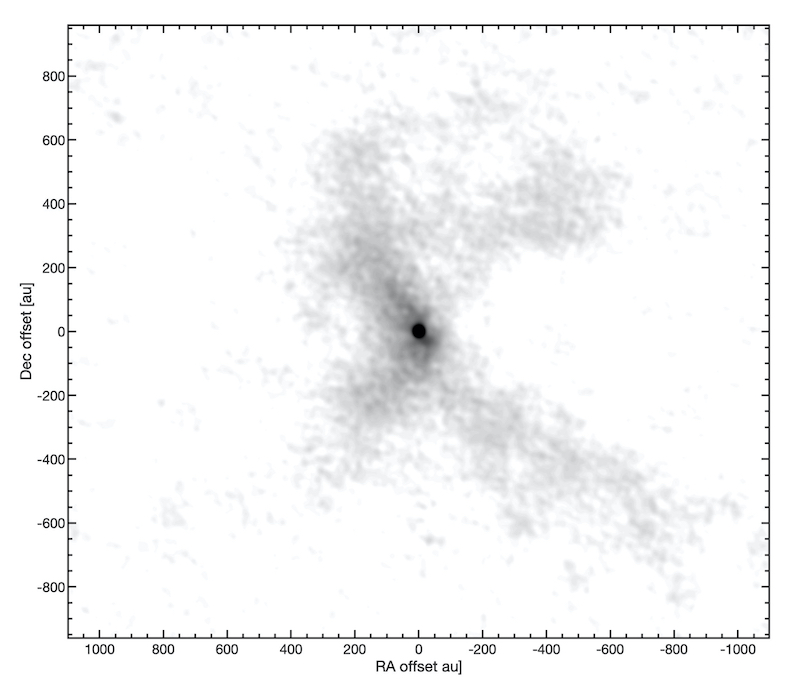} \\
   \caption{Same as Fig.  \ref{fig:moment0_combined}, but continuum only and zoomed out by a factor of two.}
   \label{fig:a0}
\end{figure*}
\begin{figure*}[ht!]
   \centering
   \includegraphics[width=0.68\hsize]{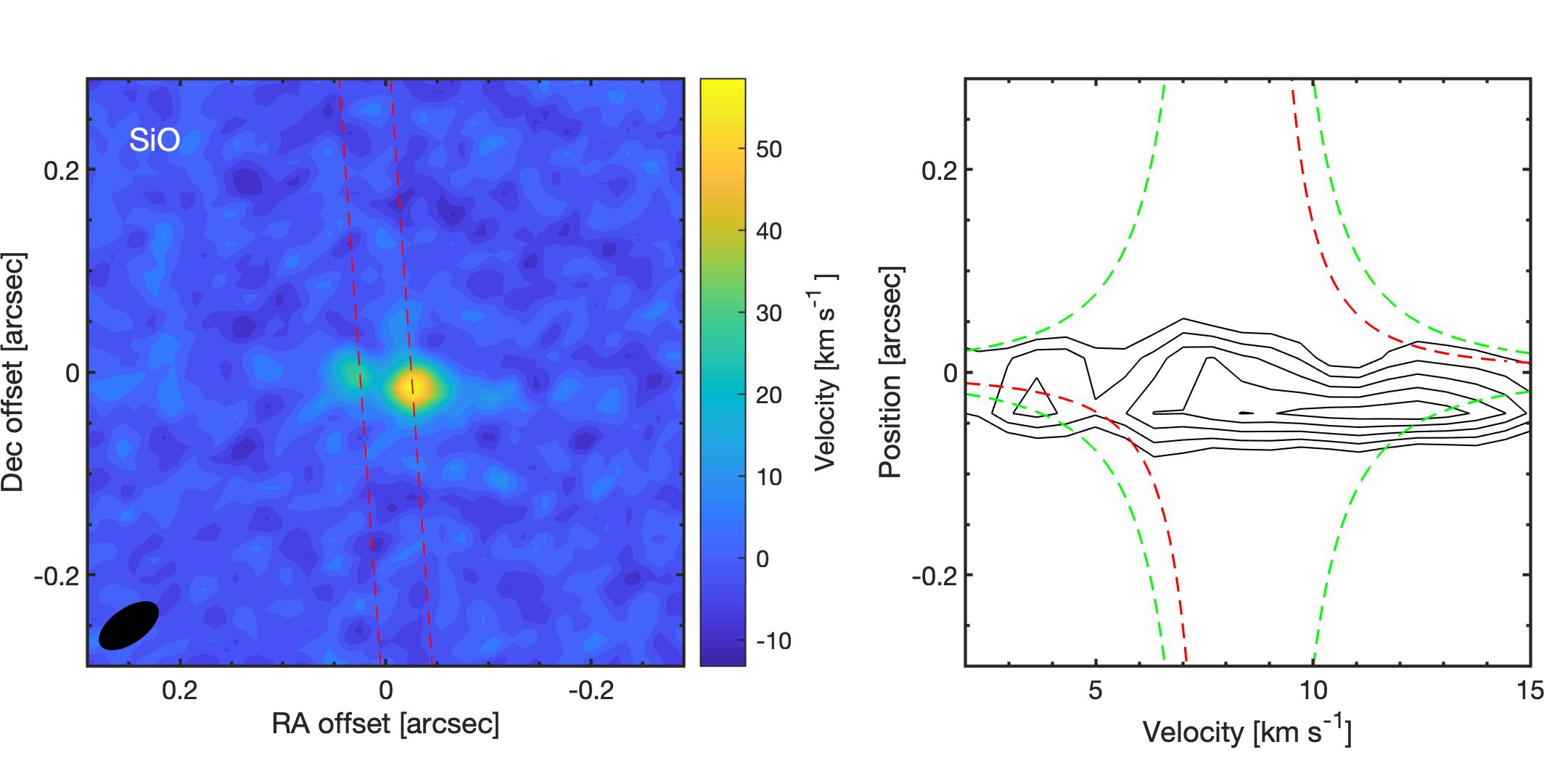} \\
   \caption{Same as Fig. \ref{fig:CH3OH_PV}, but for SiO\,(5--4). The SiO emission shows no clear evidence of rotation.
   }
   \label{fig:a1}
\end{figure*}
\begin{figure}[ht!]
   \centering
   \includegraphics[width=1.0\hsize]{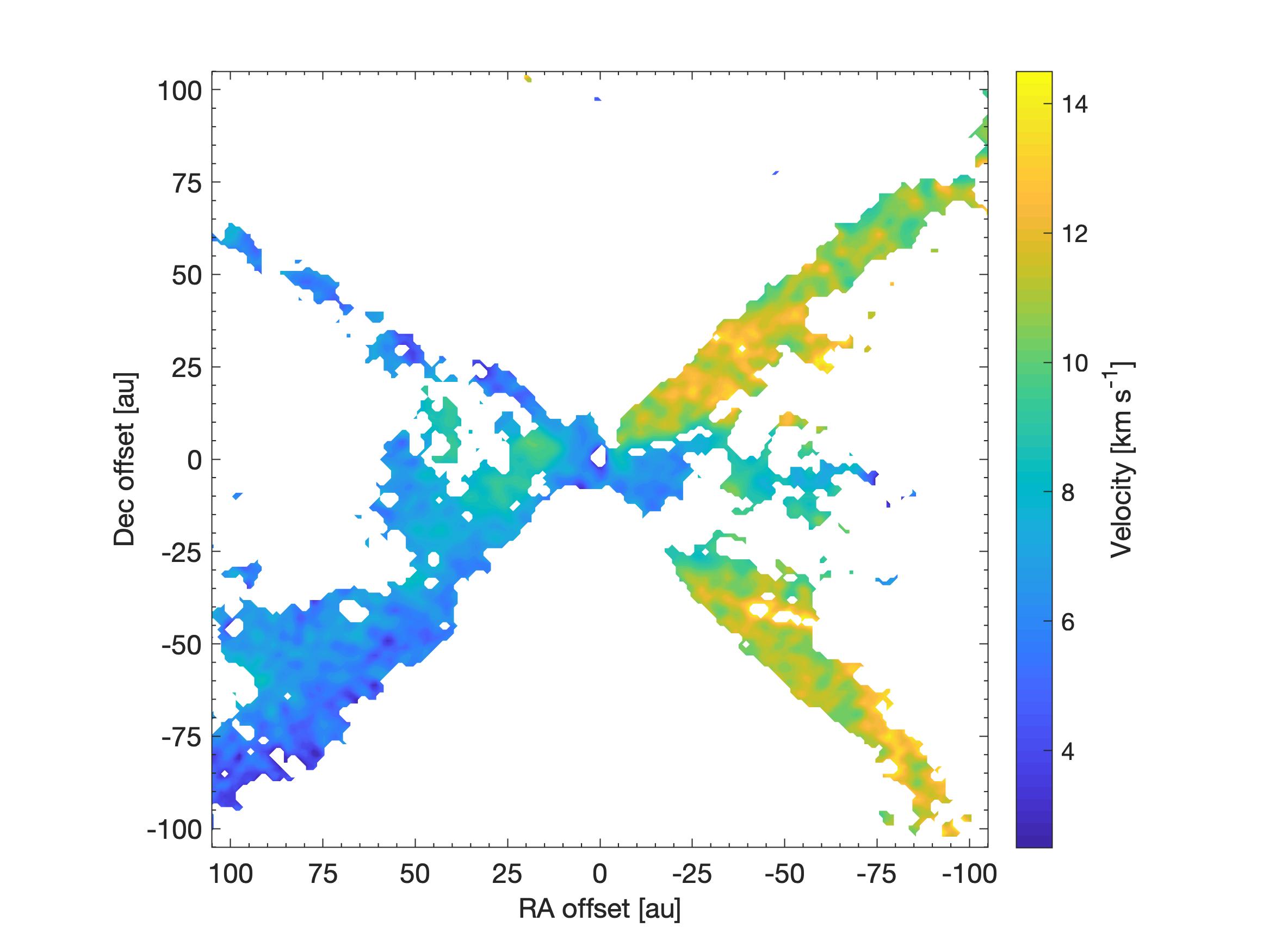} \\
   \caption{Moment 1 map for the combined \tolvco\ data.
   }
   \label{fig:a2}
\end{figure}
Figure \ref{fig:a0} presents continuum emission from Fig. \ref{fig:moment0_combined} zoomed out by a factor of two. Figure \ref{fig:a1} shows the PV diagram for the SiO\,(5--4) line detected in our long-baseline observations, but that was not presented in Fig.\ \ref{fig:CH3OH_PV}. Figure \ref{fig:a2} shows the \tolvco\ moment 1 map for the combined data set.

\end{appendix}

\end{document}